\documentclass[pra,aps,twocolumn]{revtex4}
\usepackage{amssymb}
\usepackage{latexsym}
\usepackage{graphicx}
\usepackage{boldtensors}
\newcommand{\Z}{{\mathbb Z}}

\newcommand{\R}{{\mathbb R}}

\newcommand{\LL}{{\mathbb L}}

\def\be{\begin{equation}}
\def\ee{\end{equation}}
\def\bea{\begin{eqnarray}}
\def\eea{\end{eqnarray}}
\def\Tr{{\rm \,Tr\,}}

\def\d{{\,\rm d}}

\def\bfz{{\bf 0}}
\def\a{{\bf a}}
\def\b{{\bf b}}
\def\k{{\bf k}}

\def\g{{\bf g}}

\def\q{{\bf q}}
\def\n{{\bf n}}
\def\p{{\bf p}}

\def\v{{\bf v}}
\def\x{{\bf x}}

\def\y{{\bf y}}
\def\z{{\bf z}}

\def\P{{\bf P}}

\def\Q{{\bf Q}}
\def\X{{\bf X}}
\def\Y{{\bf Y}}

\def\h2m{\frac{\hbar^2}{2m}}

\begin{document}
\title{
\vspace{1cm} \large\bf The total momentum of quantum fluids\\}
\author{Andr\'as S\"ut\H o\\
Institute for Solid State Physics and Optics, Wigner Research Centre for Physics,
Hungarian Academy of Sciences, P. O. Box 49, H-1525 Budapest,  Hungary}
\thispagestyle{empty}
\begin{abstract}
\noindent
The probability distribution of the total momentum $\P$ is studied in $N$-particle interacting homogeneous quantum systems at positive temperatures. Using Galilean invariance we prove that in one dimension the asymptotic distribution of $\P/\sqrt{N}$ is normal at all temperatures and densities, and in two dimensions the tail distribution of $\P/\sqrt{N}$ is normal. We introduce the notion of the density matrix reduced to the center of mass, and show that its eigenvalues are $N$ times the probabilities of the different eigenvalues of $\P$. A series of results is presented for the limit of sequences of positive definite atomic probability measures, relevant for the probability distribution of both the single-particle and the total momentum. The $\P=\bfz$ ensemble is shown to be equivalent to the canonical ensemble. Through some conjectures we associate the properties of the asymptotic distribution of the total momentum with the characteristics of fluid, solid, and superfluid phases. Our main suggestion is that in interacting quantum systems above one dimension, in infinite space, the total momentum is finite with a nonzero probability at all temperatures and densities. In solids this probability is 1, and in a crystal it is distributed on a lattice. Since it is less than 1 in two dimensions, we conclude that a 2D system is always in a fluid phase; so if the hexatic phase existed classically, it would be destroyed by quantum fluctuations. For a superfluid we conjecture that the total momentum is zero with a nonzero probability and otherwise its distribution is continuous. We define a macroscopic wave function based on the density matrix reduced to the center of mass. We discuss how dissipation can give rise to a critical velocity, predict the temperature dependence of the latter, and prove that Landau's criterion cannot explain superfluidity and its breakdown
in a dissipative flow.
We also comment on the relation between superfluidity and Bose-Einstein condensation.

\vspace{2mm}\noindent PACS: 03.75.Kk, 05.30.-d, 67.10.Fj, 67.25.dj, 67.80.bd

\end{abstract}
\maketitle
\noindent
\textbf{I. INTRODUCTION}

\vspace{3pt}
Proving phase transitions in continuous space is notoriously difficult. To the author's knowledge there are only three rigorous results for classical systems, by Ruelle~\cite{Rue} about segregation in the Widom-Rowlinson model, by Lebowitz, Mazel and Presutti~\cite{LMP} about the vapor-liquid transition, and by Bowen et al.~\cite{BLRW} about fluid-solid transition, and none for interacting homogeneous quantum systems. Here we suggest the use of the total momentum for the characterization of the different thermodynamic phases of quantum systems in continuous space, at positive temperatures. A short account of this work is given elsewhere~\cite{Su0}.

At a first sight the total momentum is not a very exciting quantity. In finite volumes it is conserved only if the boundary condition is periodic. Moreover, because it is the sum of $N$ single-particle momenta, one would guess it to be typically of order $\sqrt{N}$. Yet a second thought may modify this picture. Thermodynamic phases can unambiguously be characterized only in the thermodynamic limit, so if we are interested in homogeneous systems, the use of periodic boundary conditions is not really restrictive. More important, the `typical' $\sqrt{N}$ behavior does not mean exclusivity. Indeed, in interacting quantum systems there are excitations in which a macroscopic number of particles is involved. If these `collective modes' carry a momentum, as in the case of density waves and lattice vibrations, this is, or is proportional to, the total momentum; so manifestly the total momentum can remain finite in the thermodynamic limit. The center of our interest is the probability distribution of the total momentum in thermal equilibrium.

In Section II we write down a general $N$-particle Hamiltonian, the corresponding density matrix and partition function in $d$-dimensional cubes with periodic boundary conditions. We then use Galilean invariance to sum separately over eigenstates belonging to the irreducible wave vectors, introduced in Ref.~\cite{Suto}, and over their Galilean boosts. This leads to some interesting results. The set of irreducible wave vectors is markedly different in one, two, and higher dimensions. In one dimension we can derive a central limit theorem showing that the limit of $\Q/\sqrt{N}$, where $\hbar\Q$ are the eigenvalues of the total momentum operator, is normally distributed at all temperatures and densities (Theorem II.1).  In two dimensions only the tail distribution is normal (Theorem II.2), and for higher dimensions we can prove only that $|\Q|/N=O(1/L)$ almost surely (Proposition II.3).

In two theorems of Section III we investigate the relation between the total momentum and the center of mass. In Theorem III.1 first we show that in any eigenstate of the total momentum operator the center of mass can be separated and it propagates as a free particle carrying the total momentum. This result, applied already in our earlier paper~\cite{Suto}, is intuitively obvious but its derivation needs some care. We then use this separability to define $\rho_{\rm c.m.}$, the density matrix reduced to the center of mass, and prove that the eigenvalues of $\rho_{\rm c.m.}$ are $N$ times the probabilities of the different eigenvalues of the total momentum. The integral kernel $\langle\x|\rho_{\rm c.m.}|\y\rangle$ is shown to be real nonnegative; we interpret it as the autocorrelation function of the center of mass. Theorem III.2 extends the definition of $\rho_{\rm c.m.}$ to cases when a restricted density matrix is reduced to the center of mass; e.g. this can be the projection of the density matrix to the $\Q=\bfz$ subspace.

Section IV is a brief summary of some known results for the single-particle momentum distribution. We include it to stress the structural analogy between $\rho_{\rm c.m.}$ and $\rho_1$, the one-particle reduced density matrix. The expected fraction of particles in the one-particle state of momentum $\hbar\k$ is the analog of the probability that the total momentum is $\hbar\k$, and the integral kernel $\langle\x|\rho_1|\y\rangle$ is also real nonnegative; moreover, $\langle\x|\rho_{\rm c.m.}|\x\rangle=\langle\x|\rho_1|\x\rangle=\rho$, the global density. The fundamental difference between $\rho_{\rm c.m.}$ and $\rho_1$ is due to the fact that the single-particle motion is not separable [except for the noninteracting Boltzmann gas].

Section V is about the infinite volume limit of the probability distributions introduced for the total momentum in Section II and for the single-particle momentum in Section IV. In both cases we are given a sequence $\varphi_L$ of positive definite atomic probability measures concentrated on a lattice whose spacing tends to zero as $1/L$. The distribution limit of $\varphi_L$ is a not necessarily normalized positive measure in the continuous dual space. We can compactify the latter by adding a point at infinity, and attribute the missing weight to that point. The section contains a series of propositions and theorems relating the weight at infinity to the properties of the limit of $\varphi_L$ and of its Fourier transform. We also give examples that we think to be characteristic to crystals, fluids, Bose-condensates, and superfluids.

In Section VI we show that from the point of view of thermodynamic quantities the subspace of zero total momentum is representative to the whole system: the asymptotic free energy density can be computed in the $\Q=\bfz$ restricted ensemble. With this restriction we lose information about the probability distribution of $\Q$, with a notable exception: Theorem III.2 implies that the probability of $\Q=\bfz$ in the canonical ensemble can directly be obtained from the reduction of the density matrix to the center of mass within the $\Q=\bfz$ subspace.

In Section VII we make some conjectures about the relation between the probability distribution of the total momentum and the type of the thermodynamic phase, partly anticipated in the examples of Section V. The total momentum is the property of a state of the whole system. So when we say that in an infinite fluid (gas or liquid) the total momentum can be both finite and infinite, we mean that the thermal equilibrium state is a convex combination of states with different total momenta, both finite and infinite. We conjecture that in an infinite solid the random few-particle motion freezes out, the total momentum is finite with probability one and has a purely atomic probability distribution; in crystals the probability distribution is concentrated on a lattice. [In a supersolid one would have the convex combination of an atomic and a continuous probability distribution.] This corresponds to what we found in one dimension at zero temperature~\cite{Suto}. Because we proved earlier that in two dimensions at all temperatures and densities the total momentum diverges as $\sqrt{N}$ with a nonvanishing probability, cf. Theorem II.2, we conclude that no solid phase can exist in 2D; so quantum fluctuations would destroy the hexatic phase in case it existed classically.

The most audacious among the conjectures is the one about the superfluid. Superfluidity is well understood experimentally, and there exists a fully developed hydrodynamic theory initiated by London~\cite{Lon}, Tisza~\cite{Tisza} and Landau~\cite{Lan} which, assisted by the weaponry of many-body physics, provides a satisfactory explanation to practically all experimental findings. Moreover, this theory is accessible in monographs written by top experts of the subject~\cite{Lon,NP,Grif,Leg,Kag}. However, the first-principle characterization of superfluids is a slippery problem~\cite{Kad}. A pioneering work about the $\lambda$-transition in liquid helium, based on first principles, is due to Feynman~\cite{Fey1}. Through his path integral representation of the partition function, by making a series of approximations he obtained a third order phase transition with a continuously varying specific heat, instead of a discontinuous one. His method has remained the most promising route to a proof, but until today nobody was able to accomplish the task.

Our conjecture says that below the superfluid transition temperature the total momentum of the infinite system is zero with a nonzero probability, and the state of zero momentum can be identified with the superfluid component. The normal component is the state in which the total momentum is nonzero, and the thermal equilibrium state of the infinite system is the convex combination of the two components taken with the respective probabilities. A measurement can project the system into one or the other component. This definition makes the superfluid transition reminiscent to Bose-Einstein condensation (BEC), only the 'condensation' takes place into the state of zero total [and not single-particle] momentum. The discussion of the consequences of this conjecture occupies the major part of Section VII. Our definition is motivated partly by aesthetic reasons. First, there is the appealing analogy between $\rho_{\rm c.m.}$ and $\rho_1$. Second, as the temperature goes to zero, the Gibbs state tends to the ground state which is both 100\% superfluid and of zero total momentum. Third, it follows that in the superfluid state thus defined the system really moves `as a whole', as Landau phrased it~\cite{Lan}. At rest or in a hypothetic frictionless flow obtained by Galilean boost the superfluid is in thermal equilibrium; in a realistic dissipative flow the superfluid is a non-equilibrium state. The effect of dissipation can be understood as the breakdown of thermal equilibrium in such a way that in finite volumes the density matrix becomes the convex combination of two states of equal temperatures but generated by different Hamiltonians. The existence of a critical velocity together with its temperature dependence ensues from this picture. A different view of the non-equilibrium nature of the superfluid flow at zero temperature is presented in Ref.~\cite{Wre}.

At the end of Section VII we comment on the relation between superfluidity and Bose-Einstein condensation. In two dimensions there is no BEC at positive temperatures~\cite{Hoh,Wag,BM}, but a superfluid transition takes place in trapped 2D Bose gases~\cite{Dal} and in helium monolayers~\cite{KK}. Bulk liquid helium undergoes both BEC and superfluid phase transition, and there is today a consensus that Fritz London's original idea~\cite{Lon} was correct, and the superfluid phase transition and BEC occur simultaneously. Yet even in 3D one cannot equate the superfluid with the Bose condensate: as the temperature goes to zero, the whole system becomes superfluid while the Bose condensate contains less than $10\%$ of the particles~\cite{PO,AZ,SS,Gly,Cep,Mor}. The condensate wave function, obtained from $\rho_1$, is generally accepted to be the macroscopic wave function associated with the superfluid. However, a macroscopic wave function derived from $\rho_{\rm c.m.}$ can play the same role even better: because the superfluid fraction, as we define it, tends correctly to 1 with the temperature going to zero, one can avoid the embarrassing question, how less than 10\% can represent 100\%. Meanwhile, we do not challenge the consensus. In three dimensions BEC and the superfluid transition may well be simultaneous. Also, there is nothing contradictory in supposing that the total momentum and a macroscopic number of single-particle momenta vanish simultaneously. The advantage of our proposal is that it points to a common cause of superfluidity in two and three dimensions.

In Section VIII we discuss Landau's criterion for the breakdown of superfluidity. Doubts about its validity were raised already earlier, see e.g. in Refs.~\cite{BaPe,Kad}. Here we analyze Landau's original publication~\cite{Lan} and find
that his argument does not apply to a dissipative flow.
The paper ends with a Summary.

\vspace{5pt}
\noindent
{\bf II. DISTRIBUTION OF THE TOTAL MOMENTUM}

\vspace{3pt}
The setting is the same as in Ref.~\cite{Suto}.
\be
H=\frac{1}{2m}\sum_{j=1}^N \p_j^2+U_\Lambda(\x_1,\ldots,\x_N)
\ee
is the energy operator of $N$ interacting particles in a $d$-dimensional cube $\Lambda$ of side length $L$, defined with periodic boundary conditions; $U_\Lambda$ is the periodized potential energy. Stability is supposed, and hard-core interactions are allowed, with the suitable modification of the domain of $H$. $H$ is either unrestricted [Boltzmann statistics] or restricted to the symmetric or antisymmetric subspace. The common eigenstates of $H$ and the total momentum operator
\be
\P=\sum_j\p_j\equiv -i\hbar\sum_j \partial/\partial\x_j
\ee
are denoted by $\psi_{\Q,n}$; the respective eigenvalues are $E_{\Q,n}$, and $\hbar\Q$, where
\be
\Q\in\Lambda^*=(2\pi/L)\Z^d.
\ee
The energies are ordered, $E_{\Q,n}\leq E_{\Q,n+1}\quad (n\geq 0)$. The ground state and its energy are $\psi_{\bfz,0}$ and $E_{\bfz,0}$, respectively. With the notation
\be\label{projection-Qn}
\quad \Pi_{\Q,n}=|\psi_{\Q,n}\rangle\langle\psi_{\Q,n}|
\ee
the unnormalized density matrix is
\be
e^{-\beta H}=\sum_{\Q\in\Lambda^*}\sum_{n=0}^\infty e^{-\beta E_{\Q,n}}\Pi_{\Q,n}.
\ee
Introducing the set of irreducible wave vectors
\be\label{lambdairred}
\Lambda^*_{\rm irred}=\{\q\in\Lambda^*: -\pi N/L<q_i\leq\pi N/L\ \mbox{all $i$}\},
\ee
any $\Q\in\Lambda^*$ can uniquely be written as $\q+N\k$, where $\q\in\Lambda^*_{\rm irred}$ and $\k\in\Lambda^*$. Then,
\be\label{ebetaH}
e^{-\beta H}=\sum_{\q\in\Lambda^*_{\rm irred}}\sum_{\k\in\Lambda^*}\sum_{n=0}^\infty e^{-\beta E_{\q+N\k,n}}\Pi_{\q+N\k,n}.
\ee
By Galilean invariance~\cite{Suto},
\be\label{Galilei}
E_{\q+N\k,n}=E_{\q,n}+(\hbar^2/2m)[N\k^2+2\k\cdot\q].
\ee
Substituting this into Eq.~(\ref{ebetaH}),
\bea\label{DM}
e^{-\beta H}&=&\sum_{\q\in\Lambda^*_{\rm irred}}\sum_{n=0}^\infty e^{-\beta E_{\q,n}}
\nonumber\\
&\times&\sum_{\k\in\Lambda^*}e^{-(\beta\hbar^2/2m)[N\k^2+2\k\cdot\q]} \Pi_{\q+N\k,n}.
\eea
In what follows, we fix the number density $\rho$; then $N=\rho L^d$. With $\lambda_\beta=\sqrt{2\pi\beta\hbar^2/m}$, the partition function reads
\be\label{Gauss}
Z=\Tr e^{-\beta H}=\sum_{\q\in\Lambda^*_{\rm irred}}\sum_{n=0}^\infty e^{-\beta E_{\q,n}}
\sum_{\k\in\Lambda^*}e^{-\frac{\lambda_\beta^2}{4\pi}[N\k^2+2\k\cdot\q]}.
\ee
Let
\be\label{Zirred0}
Z_{\rm irred}=\sum_{\q\in\Lambda^*_{\rm irred}}\sum_{n=0}^\infty e^{-\beta E_{\q,n}},
\ee
the $\k=\bfz$ term of $Z$. Then $Z/Z_{\rm irred}$ is the average in $\Lambda^*_{\rm irred}$ of the sum over $\k$,
\be
\frac{Z}{Z_{\rm irred}}=\left\langle \sum_{\k\in\Lambda^*}e^{-\frac{\lambda_\beta^2}{4\pi}[N\k^2+2\k\cdot\q]}\right\rangle_{\rm irred}.
\ee
First we bound this quantity.

\vspace{5pt}
\noindent
{\bf Proposition II.1.}
\be\label{up-low}
\max\left\{1,\left[\frac{L^{1-d/2}}{\lambda_\beta\sqrt{\rho}}-1\right]^d\right\} \leq\frac{Z}{Z_{\rm irred}} \leq \left[\frac{L^{1-d/2}}{\lambda_\beta\sqrt{\rho}}+3\right]^d.
\ee

\noindent
{\em Proof.} The sum over $\k$ in Eq.~(\ref{Gauss}) can be factorized. Replacing the $i$th component of $\k$ by $2\pi j/L$,
\bea\label{ksum}
\sum_{\k\in\Lambda^*}e^{\frac{-\lambda_\beta^2}{4\pi}[N\k^2+2\k\cdot\q]} =\prod_{i=1}^d\left[\sum_{j=-\infty}^\infty e^{-\frac{\pi\lambda_\beta^2 Nj^2}{L^2}}\cosh\frac{\lambda_\beta^2 q_i j}{L}\right]\nonumber\\
=\prod_{i=1}^d\left[1+2\sum_{j=1}^\infty e^{-\frac{\pi\lambda_\beta^2 Nj^2}{L^2}}\cosh\frac{\lambda_\beta^2 q_i j}{L}\right].\nonumber\\
\phantom{a}
\eea
For $\q\in\Lambda^*_{\rm irred}$ and $j>0$,
\be\label{low-cosh-up}
1\leq \cosh\frac{\lambda_\beta^2 q_i j}{L}\leq \exp\frac{\pi\lambda_\beta^2 Nj}{L^2}.
\ee
Equation~(\ref{up-low}) follows from the upper and lower bounds
\be
\sum_{j=1}^\infty e^{-\frac{\pi\lambda_\beta^2 N}{L^2}j(j-1)}<1+\int_0^\infty e^{-\frac{\pi\lambda_\beta^2 N}{L^2}x^2}\d x
\ee
and
\be
\sum_{j=1}^\infty e^{-\frac{\pi\lambda_\beta^2 N}{L^2}j^2}>\max\left\{0,\int_0^\infty e^{-\frac{\pi\lambda_\beta^2 N}{L^2}x^2}\d x-1\right\}
\ee
with
$$
\int_0^\infty e^{-\frac{\pi\lambda_\beta^2 N}{L^2}x^2}\d x=\frac{L^{1-d/2}}{2\lambda_\beta\sqrt{\rho}}.\quad\Box
$$

Let
\be\label{nuQ}
\nu_\Q=Z^{-1}\sum_{n=0}^\infty e^{-\beta E_{\Q,n}} =Z^{-1}\sum_{n=0}^\infty e^{-\beta E_{\q,n}}e^{-\frac{\lambda_\beta^2}{4\pi}[N\k^2+2\k\cdot\q]},
\ee
the probability that in thermal equilibrium the system is in the state $\sum_n e^{-\beta E_{\Q,n}}\Pi_{\Q,n}/\sum_n e^{-\beta E_{\Q,n}}$ with $\hbar\Q=\hbar(\q+N\k)$.
By definition, $\sum_{\Q\in\Lambda^*}\nu_\Q=1$, meaning that in finite volumes the total momentum is finite with probability 1. As we shall see, this is not the case in infinite volume. Let us introduce also the probability distribution function
\be
\Gamma^N_{L}(\kappa)=\sum_{\|\Q\|\leq \kappa}\nu_\Q
\ee
where $\|\Q\|=\max_{1\leq i\leq d}|Q_i|$. In words, $\Gamma^N_{L}(\kappa)$ is the probability that the modulus of each component of the total momentum is less than or equal to $\hbar\kappa$. Then
\be
\frac{Z_{\rm irred}}{Z}=\sum_{\q\in\Lambda^*_{\rm irred}}\nu_\Q\approx\Gamma^N_{L}\left(\pi N/L\right),
\ee
cf. Eq.~(\ref{lambdairred}),
and Eq.~(\ref{up-low}) can be put in the form
\be\label{2bounds}
\frac{1}{\left[\frac{L^{1-d/2}}{\lambda_\beta\sqrt{\rho}}+3\right]^{d}} \leq\sum_{\q\in\Lambda^*_{\rm irred}}\nu_\Q \leq\min \left\{1,\frac{1}{\left[\frac{L^{1-d/2}}{\lambda_\beta\sqrt{\rho}}-1\right]^{d}}\right\}.
\ee
By taking the limit $L\to\infty$, a first estimate about the asymptotic behavior of $\Gamma^N_L$ can be obtained.

\vspace{5pt}
\noindent
{\bf Proposition II.2.} For any $\beta$ and $\rho$,
\begin{itemize}
\item{if $d=1$,}
\be\label{irred1}
\lim_{L\to\infty}\Gamma^N_{L}(\pi\rho)=0;
\ee
\item{if $d=2$,}
\be\label{irred2}
\lim_{L\to\infty}\Gamma^N_L(\pi\rho L)\geq \frac{\lambda_\beta^2\rho}{\left(1+3\lambda_\beta\sqrt{\rho}\right)^2};
\ee
\item{if $d\geq 3$,}
\be\label{irred>2}
\lim_{L\to\infty}\Gamma^N_L\left(\pi\rho L^{d-1}\right)\geq 1/3^d.
\ee
\end{itemize}

\vspace{3pt}
\noindent
{\em Proof.} For $d=1$ the upper bound, for $d\geq 2$ the lower bound in Eq.~(\ref{2bounds}) provides the result.\ $\Box$

\vspace{3pt}
Because $\Q$ is the sum of $N$ single-particle wave vectors, normal fluctuations of the total momentum about the zero expectation value would mean that the absolute value of each component of $\Q$ is typically of order $\sqrt{N}$. Let $J$ be any positive integer and  consider the probability
\be
F_{2J-1}\equiv{\rm Prob}\left\{|Q_1|>\frac{(2J-1)\pi N}{L}\right\}=\sum_{|Q_1|>\frac{(2J-1)\pi N}{L}}\nu_\Q.
\ee
From Eqs.~(\ref{ksum}) and (\ref{nuQ}),
\bea\label{prob}
F_{2J-1}
&=&\frac{2}{Z}\sum_{n=0}^\infty\sum_{\q\in\Lambda^*_{\rm irred}}e^{-\beta E_{\q,n}}
\left[\sum_{j=J}^\infty e^{-\frac{\pi\lambda_\beta^2 Nj^2}{L^2}}
\cosh\frac{\lambda_\beta^2 q_1 j}{L}\right]\nonumber\\
&\times&\prod_{i=2}^{d-1}\left[\sum_{j=-\infty}^\infty e^{-\frac{\pi\lambda_\beta^2 Nj^2}{L^2}}
\cosh\frac{\lambda_\beta^2 q_i j}{L}\right].
\eea

\vspace{5pt}
\noindent
{\bf Theorem II.1.} In one dimension the asymptotic distribution of $Q/\sqrt{N}$ is normal at all temperatures and densities. Namely,
\be\label{CLT1D}
\lim_{N\to\infty}{\rm Prob}\left\{\frac{|Q|}{\rho\sqrt{N}}\leq x\right\}
=\frac{\lambda_\beta\rho}{\pi}\int_0^x e^{-\frac{\lambda_\beta^2\rho^2}{4\pi}y^2}\d y.
\ee

\vspace{3pt}
\noindent
{\em Proof.}
From Eq.~(\ref{prob}) we subtract the same equation written for $J+1$:
\be
F_{2J-1}-F_{2J+1}=\frac{2}{Z}\sum_{n=0}^\infty\sum_{q\in\Lambda^*_{\rm irred}}e^{-\beta E_{q,n}} e^{-\frac{\pi\lambda_\beta^2 \rho J^2}{L}}\cosh\frac{\lambda_\beta^2 q J}{L}.
\ee
Because $|q|\leq\pi\rho$ if $q\in\Lambda^*_{\rm irred}$, summing over $J$ from 1 to $M$ and using the bounds
\be
1\leq \cosh\left(\lambda_\beta^2 qJ/L\right)\leq\cosh\left(\pi\lambda_\beta^2\rho M/L\right)
\ee
we find
\bea
\frac{2Z_{\rm irred}}{Z}\sum_{J=1}^Me^{-\pi\lambda_\beta^2\rho J^2/L}\leq F_{1}-F_{2M+1}\nonumber\\ \leq\cosh\left(\pi\lambda_\beta^2\rho M/L\right)\frac{2Z_{\rm irred}}{Z}\sum_{J=1}^Me^{-\pi\lambda_\beta^2\rho J^2/L}.
\eea
Now
\be
F_1-F_{2M+1}=\Gamma^N_L\left[(2M+1)\pi\rho\right]-\Gamma^N_L(\pi\rho),
\ee
where $\Gamma^N_L(\pi\rho)\to 0$, c.f. Eq.~(\ref{irred1}). Setting $M=x\sqrt{N}$, $\cosh\left(\pi\lambda_\beta^2\rho M/L\right)\to 1$. Furthermore,
\be
\lim_{N\to\infty}\frac{1}{\sqrt{N}}\sum_{J=1}^{x\sqrt{N}}e^{-\pi\lambda_\beta^2\rho J^2/L} =\int_0^x e^{-\pi\lambda_\beta^2\rho^2 y^2}\d y.
\ee
With the bounds~(\ref{2bounds}),
\be
\lim_{N\to\infty}\Gamma^N_L\left[(2x\sqrt{N}+1)\pi\rho\right] =2\lambda_\beta\rho\int_0^x e^{-\pi\lambda_\beta^2\rho^2 y^2}\d y
\ee
which is equivalent to Eq.~(\ref{CLT1D}).\ $\Box$

\vspace{5pt}
The above result obviously implies that for any sequence $a_L\to\infty$,
\be\label{1D-aL}
\lim_{L\to\infty} \Gamma^N_{L}\left(\rho \sqrt{N}a_L\right)=1,\ \lim_{L\to\infty} \Gamma^N_{L}\left(\rho \sqrt{N}/a_L\right)=0.
\ee

\vspace{5pt}
\noindent
{\bf Theorem II.2.} In two dimensions the asymptotic tail distribution of $|Q_1|/\sqrt{N}$ is normal at all temperatures and densities. More precisely, for any $J\geq 1$,
\bea\label{low-2D-up}
2f\left(\lambda_\beta\sqrt{\rho}\right)\sum_{j=J}^\infty e^{-\pi\lambda_\beta^2\rho j^2} &\leq&\lim_{N\to\infty} {\rm Prob}\left\{\frac{|Q_1|}{\sqrt{\rho N}}>(2J-1)\pi\right\}\nonumber\\
&\leq& 2f\left(\lambda_\beta\sqrt{\rho}\right)\sum_{j=J}^\infty e^{-\pi\lambda_\beta^2\rho j(j-1)}
\eea
where
\be
f\left(\lambda_\beta\sqrt{\rho}\right)=\lim_{N\to\infty}\frac{Z_{\rm irred}}{Z} \left\langle\sum_{j=-\infty}^\infty e^{-\pi\lambda_\beta^2\rho j^2}\cosh\frac{\lambda_\beta^2 q_2 j}{L}\right\rangle_{\rm irred}.
\ee

\vspace{3pt}
\noindent
{\em Proof.} We apply Eq.~(\ref{prob}) with the lower and upper bounds obtained from Eq.~(\ref{low-cosh-up}). $f\left(\lambda_\beta\sqrt{\rho}\right)>0$ follows from Eq.~(\ref{irred2}). $\lambda_\beta\sqrt{\rho}$ is the only relevant dimensionless quantity, whence the dependence of $f$ on it. For $J$ large enough the upper bound in Eq.~(\ref{low-2D-up}) is less than 1. Thus, the limit of the probability is between 0 and 1 and tends to zero as $e^{-{\rm const.}\times J^2}$ with $J$ going to infinity.
\ $\Box$

Equations~(\ref{2bounds}) and (\ref{prob}) have only a much weaker implication for $d\geq 3$.

\vspace{3pt}
\noindent
{\bf Proposition II.3.} If $d\geq 3$ then for any $\beta$ and $\rho$,
\be
\lim_{L\to\infty} \Gamma^N_{L}(3\pi N/L)=1.
\ee

\vspace{3pt}
\noindent
{\em Proof.}
With $J=2$ the first sum over $j$ in Eq.~(\ref{prob}) goes to zero while the remaining sums go to 1. So $F_3\to 0$ which gives the result.\ $\Box$

\vspace{3pt}
Above two dimensions we do not obtain the expected $\sqrt{N}$ instead of $N^{1-1/d}$ because for $d\geq 3$ any $\Q$ of order $\sqrt{N}$ is deeply inside $\Lambda^*_{\rm irred}$, and one should conceive a proof which takes the interaction into account. An idea for such a proof is as follows. The result for one and two dimensions implies that the thermal average $\langle\P^2\rangle=O(N)$. Because $\langle\p_j^2\rangle$ is finite, $\langle\p_i\cdot\p_j\rangle=O(1/N)$. The argument can work in the other direction. If one could prove that $\langle\p_i\cdot\p_j\rangle=O(1/N)$, one could conclude that $\langle\P^2\rangle=O(N)$. If the scattering length of the interaction is finite, the momentum of each particle is correlated only with that of other particles within scattering distance.
But for fixed $\x_i$, $\x_j$ can be anywhere in $\Lambda$, therefore $\langle\p_i\cdot\p_j\rangle=O(L^{-d})=O(1/N)$. Note that irrespective of an eventual improvement of the result for $d\geq 3$, in all dimensions the velocity $\P/Nm$ of the center of mass tends to zero with probability 1 as $L$ goes to infinity.

\vspace{5pt}
\noindent
{\bf III. TOTAL MOMENTUM AND CENTER OF MASS}

Let $\overline{\x}$ denote the arithmetic mean (center of mass) of the vectors $\x_1,\ldots,\x_N$. It is easy to check that
for any differentiable function $f(\x_1,\ldots,\x_N)$,
\be
(\overline{\x}\cdot\P-\P\cdot\overline{\x})f=id\hbar f,
\ee
that is, the total momentum is canonically conjugate to the center of mass. A more interesting relation between the two is as follows.

\vspace{5pt}
\noindent
{\bf Theorem III.1.} $\{N\nu_\Q,L^{-\frac{d}{2}}e^{i\Q\cdot\overline{\x}}\}_{\Q\in\Lambda^*}$ are the eigenvalues and corresponding eigenfunctions of a reduced density matrix, where the reduction is onto the center of mass. In Boltzmann and Bose statistics, $\nu_\Q$ is a positive definite function of $\Q$ on $\Lambda^*$, i.e., the $n\times n$ matrix $[\nu_{\Q_i-\Q_j}]_{ij}$ is positive semidefinite for any $n$ and any $\Q_1,\ldots,\Q_n\in\Lambda^*$. In particular,
\be\label{qmax=0}
\nu_\bfz\geq\nu_\Q,\quad\mbox{any $\Q\in\Lambda^*$}.
\ee

\noindent
{\em Proof.} (i) Separability of the center of mass.
Any eigenstate of the total momentum, hence, also $\psi_{\Q,n}$ can be written as
\be
\psi_{\Q,n}(\x_1,\ldots,\x_N)=e^{i\Q\cdot\overline{\x}}\phi(\X'),
\ee
where
\be\label{xbar-xprime}
\overline{\x}=\frac{1}{N}\sum_{j=1}^N\x_j, \quad \X'=(\x'_2,\ldots,\x'_N),\quad \x'_j=\x_j-\x_1
\ee
[so $\P\phi=0$]. Indeed, for any $\a\in\R^d$,
\be
e^{-\a\cdot\sum_j\frac{\partial}{\partial\x_j}}\psi_{\Q,n}(\x_1,\ldots,\x_N) =e^{-i\a\cdot\Q}\psi_{\Q,n}(\x_1,\ldots,\x_N),
\ee
and also
\be
e^{-\a\cdot\sum_j\frac{\partial}{\partial\x_j}}\psi_{\Q,n}(\x_1,\ldots,\x_N) =\psi_{\Q,n}(\x_1-\a,\ldots,\x_N-\a).
\ee
Comparing the two,
\be\label{shift}
\psi_{\Q,n}(\x_1,\ldots,\x_N)=e^{i\Q\cdot\a}\psi_{\Q,n}(\x_1-\a,\ldots,\x_N-\a).
\ee
From Eq.~(\ref{shift}) with $\a=\x_1$ and
\be
\x_1=\overline{\x}-N^{-1}\sum_{j=2}^N\x'_j
\ee
we find
\be\label{psix-separated}
\psi_{\Q,n}(\x_1,\ldots,\x_N)=e^{i\Q\cdot\overline{\x}} e^{-(i/N)\Q\cdot\sum_{j=2}^N\x'_j}\psi_{\Q,n}(\bfz,\X').
\ee
Introduce
\be\label{XY}
\X=(\x_1,\ldots,\x_N),\quad \Y=(\y_1,\ldots,\y_N)
\ee
with
\be\label{xjyj}
\x_j=\overline{\x}+\x'_j-\frac{1}{N}\sum_{i=2}^N\x'_i, \quad \y_j=\overline{\y}+\x'_j-\frac{1}{N}\sum_{i=2}^N\x'_i,
\ee
where $\overline{\x},\overline{\y},\x'_2,\ldots,\x'_N$ are arbitrary and $\x'_1=\bfz$. Then Eq.~(\ref{xbar-xprime}) holds true, so $\overline{\x}$ is the center of mass of $\{\x_j\}$, and one can check that $\overline{\y}$ is the center of mass of $\{\y_j\}$. Equation~(\ref{psix-separated}) is valid and, because $\y_j-\x_j=\overline{\y}-\overline{\x}$ independently of $j$, one can apply Eq.~(\ref{shift}) with $\a=\overline{\x}-\overline{\y}$, yielding
\bea
\psi_{\Q,n}(\Y)&=&e^{i\Q\cdot(\overline{\y}-\overline{\x})}\psi_{\Q,n}(\X)\nonumber\\
&=&e^{i\Q\cdot\overline{\y}} e^{-(i/N)\Q\cdot\sum_{j=2}^N\x'_j}\psi_{\Q,n}(\bfz,\X').
\eea
As a result,
\be
\psi_{\Q,n}(\X)\psi^*_{\Q,n}(\Y) =e^{i\Q\cdot(\overline{\x}-\overline{\y})}\left|\psi_{\Q,n}(\bfz,\X')\right|^2.
\ee
Because of the periodic boundary condition, integration over $\X'$ in $\Lambda^{N-1}$ gives the same result as integration over $\x_2,\ldots,\x_N$. Therefore, using again Eq.~(\ref{shift}),
\be
\int\left|\psi_{\Q,n}(\bfz,\X')\right|^2\d\X' =\int\left|\psi_{\Q,n}(\X)\right|^2\d\x_2\cdots\d\x_N=L^{-d}
\ee
independently of $\x_1$. Now we can define $\rho_{\rm c.m.}$, the density matrix reduced to the center of mass~\cite{Het}, through a partial trace over $\X'$:
\bea\label{rhocm_kernel}
\langle\overline{\x}|\rho_{\rm c.m.}|\overline{\y}\rangle&=&NZ^{-1}\sum_{\Q,n}e^{-\beta E_{\Q,n}}\int\psi_{\Q,n}(\X)\psi^*_{\Q,n}(\Y)\d\X'\nonumber\\
&=&\rho\sum_{\Q\in\Lambda^*}\nu_\Q e^{i\Q\cdot(\overline{\x}-\overline{\y})}.
\eea
It is seen that
\be
\langle\overline{\x}|\rho_{\rm c.m.}|\overline{\y}\rangle\leq\langle\overline{\x}|\rho_{\rm c.m.}|\overline{\x}\rangle=\rho\sum_{\Q\in\Lambda^*}\nu_\Q=\rho.
\ee
Fourier transformation of Eq.~(\ref{rhocm_kernel}) in $\overline{\x}$ and $\overline{\y}$ shows that $\rho_{\rm c.m.}$ is diagonal in momentum representation,
\be
\langle\Q|\rho_{\rm c.m.}|\Q'\rangle=N\delta_{\Q,\Q'}\nu_\Q,
\ee
so its eigenvalues are $N\nu_\Q$, its spectral resolution reads
\be\label{rhocm}
\rho_{\rm c.m.}=N\sum_{\Q\in\Lambda^*}\nu_\Q|\Q\rangle\langle\Q|
\ee
with $\langle\overline{\x}|\Q\rangle=L^{-\frac{d}{2}}e^{i\Q\cdot\overline{\x}}$.

\vspace{3pt}
\noindent
(ii) Positive definiteness of $\nu_\Q$. We prove that
\be\label{rhocm>0}
\langle\overline{\x}|\rho_{\rm c.m.}|\overline{\y}\rangle\geq 0.
\ee
$\rho_{\rm c.m.}$ can also be obtained from the integral kernel of $e^{-\beta H}$, without passing through the spectral resolution of the latter. With the same $\X$ and $\Y$ as above, in the Boltzmann case we obtain
\be
\langle\overline{\x}|\rho_{\rm c.m.}|\overline{\y}\rangle = \frac{N}{Z}\int_{\Lambda^{N-1}}\d\X' \left\langle\X|e^{-\beta H}|\Y\right\rangle;
\ee
for Bose statistics
\be
\langle\overline{\x}|\rho_{\rm c.m.}|\overline{\y}\rangle = \frac{1}{Z(N-1)!}\sum_{G\in S_N}\int_{\Lambda^{N-1}}\d\X' \left\langle G\X|e^{-\beta H}|\Y\right\rangle
\ee
where $S_N$ is the symmetric group of $N$ elements, and
\be
G\X=(\x_{G(1)},\ldots,\x_{G(N)})
\ee
[here $H$ is unrestricted; restriction to the symmetric subspace is explicitly taken into account]. Equation~(\ref{rhocm>0}) follows from
\be\label{kernel}
\langle\X|e^{-\beta H}|\Y\rangle\geq 0
\ee
which can be traced back through the Trotter formula to
\bea\label{positivity}
\langle\x|e^{-\alpha'\p_j^2}|\y\rangle
=L^{-d}\sum_{\k\in\Lambda^*}e^{-\alpha |\k|^2}e^{i\k\cdot(\x-\y)}\nonumber\\
=\left[\frac{1}{4\pi\alpha}\right]^{\frac{d}{2}}e^{-\frac{|\x-\y|^2}{4\alpha}} \sum_{\n\in\Z^d}e^{-\frac{L^2}{4\alpha}|\n|^2-\frac{L}{2\alpha}\n\cdot(\x-\y)}>0,
\eea
obtained via the Poisson summation formula [$\alpha=\alpha'\hbar^2/2m$]. Therefore,
\bea\label{nQ-Fourier}
N\nu_\Q&=&\langle\Q|\rho_{\rm c.m.}|\Q\rangle
=\frac{1}{L^d}\int_{\Lambda^2}\langle\overline{\x}|\rho_{\rm c.m.}|\overline{\y}\rangle e^{i\Q\cdot(\overline{\y}-\overline{\x})}\d\overline{\x}\d\overline{\y}\nonumber\\
&=&\int_\Lambda\langle\bfz|\rho_{\rm c.m.}|{\z}\rangle e^{i\Q\cdot\z}\d{\z}
\leq N\nu_\bfz.
\eea
Being the Fourier transform of an integrable positive function, $\nu_\Q$ is a positive definite function (a function of positive type~\cite{RS}). $\Box$

Because $\nu_\Q\geq 0$, by Eqs.~(\ref{rhocm}) and (\ref{rhocm_kernel}) $\rho_{\rm c.m.}$ is a positive operator, and its kernel is also a positive definite function: the $n\times n$ matrix $[\langle\x_i|\rho_{\rm c.m.}|\x_j\rangle]_{ij}$ is positive semidefinite for any $n$ and $\x_1,\ldots,\x_n\in\Lambda$. Clearly, $\langle\x|\rho_{\rm c.m.}|\y\rangle=\langle\y|\rho_{\rm c.m.}|\x\rangle$ or $\langle\bfz|\rho_{\rm c.m.}|{\z}\rangle=\langle\bfz|\rho_{\rm c.m.}|\!-\z\rangle$. One can interpret $\rho^{-1}\langle\x|\rho_{\rm c.m.}|\y\rangle$ as the autocorrelation function of the center of mass.

Note that Eq.~(\ref{kernel}) remains valid in the subspace of zero total momentum. We prove a more general result. Introduce
\be
\Pi_\Q=\sum_{n=0}^\infty\Pi_{\Q,n}.
\ee
By definition~(\ref{projection-Qn}), $\sum_{\Q\in\Lambda^*}\Pi_\Q=I_N$, the identity operator in $L^2(\Lambda^N)$.

\vspace{5pt}
\noindent
{\bf Theorem III.2.} Let $\LL^*$ be any subgroup of $\Lambda^*$, and let $\phi=\{\phi_\Q\}$ be a positive definite function on $\LL^*$. Then
\be\label{phiPi}
\left\langle\X\left|\sum_{\Q\in\LL^*}\phi_\Q\Pi_\Q e^{-\beta H}\right|\Y\right\rangle\geq 0.
\ee
Moreover, if we define $\rho^\phi_{\rm c.m.}$ by its integral kernel as
\be
\langle\overline{\x}|\rho^\phi_{\rm c.m.}|\overline{\y}\rangle =NZ^{-1}\int\left\langle\X\left|\sum_{\Q\in\LL^*}\phi_\Q\Pi_\Q e^{-\beta H}\right|\Y\right\rangle\d\X'
\ee
where $\X$ and $\Y$ are given by Eqs.~(\ref{XY}), (\ref{xjyj}), then
\be\label{rhocm-phi}
\langle\overline{\x}|\rho^\phi_{\rm c.m.}|\overline{\y}\rangle =\rho\sum_{\Q\in\LL^*}\phi_\Q\nu_\Q e^{i\Q\cdot(\overline{\x}-\overline{\y})}\geq 0.
\ee
In particular, with $\phi_\Q=\delta_{\Q,\bfz}$ we obtain
\be
\left\langle\X\left|\Pi_\bfz e^{-\beta H}\right|\Y\right\rangle\geq 0
\ee
and $\Pi_\bfz\rho_{\rm c.m.}=N\nu_\bfz|\bfz\rangle\langle\bfz|$ or
\be\label{rhocm^0}
\langle\overline{\x}|\Pi_\bfz\rho_{\rm c.m.}|\overline{\y}\rangle \equiv\rho\nu_\bfz.
\ee

\noindent
{\em Proof.} Let $\X-\a=(\x_1-\a,\ldots,\x_N-\a)$. Then
\bea\label{Q-restricted}
&\langle&\!\!\!\!\!\X|\Pi_\Q e^{-\beta H}|\Y\rangle =\sum_n e^{-\beta E_{\Q,n}} \psi_{\Q,n}\left(\X\right)\psi_{\Q,n}^*\left(\Y\right)\nonumber\\
&=&\sum_{\Q'\in\Lambda^*}\delta_{\Q',\Q}\sum_n e^{-\beta E_{\Q',n}}\psi_{\Q',n}\left(\X\right)\psi_{\Q',n}^*\left(\Y\right)\nonumber\\
&=&L^{-d}\int_\Lambda\d\a\,e^{i\Q\cdot\a}\sum_{\Q',n} e^{-\beta E_{\Q',n}}\psi_{\Q',n}\left(\X-\a\right)\psi_{\Q',n}^*\left(\Y\right)\nonumber\\
&=&L^{-d}\int_\Lambda\d\a\,e^{i\Q\cdot\a}\left\langle \X-\a|e^{-\beta H}|\Y\right\rangle.
\eea
Multiplying by $\phi_\Q$ and summing over $\Q$ in $\LL^*$ we find
$$\sum_{\Q\in\LL^*}\phi_\Q e^{i\Q\cdot\a}\geq 0$$
(in distribution sense) which together with Eq.~(\ref{kernel}) gives Eq.~(\ref{phiPi}). Above we used Eq.~(\ref{shift}). It is seen that $\langle\X|\Pi_\Q e^{-\beta H}|\Y\rangle$ is also a positive definite function of $\Q$. Reduction of $\Pi_\Q e^{-\beta H}$ to the center of mass provides directly $\nu_\Q|\Q\rangle\langle\Q|$, from which Eq.~(\ref{rhocm-phi}) follows. $\Box$

\vspace{5pt}
\noindent
{\bf IV. DISTRIBUTION OF THE SINGLE-PARTICLE MOMENTUM}

\vspace{3pt}
For comparison, here we collect some known results. The occupation-number operator $N_\k$ for the one-particle state $|\k\rangle$ [$\k\in\Lambda^*$] is a sum of projectors,
\be
N_\k=\sum_{j=1}^N N_{\k,j},\quad N_{\k,j}=I_{j-1}\otimes|\k\rangle\langle\k|\otimes I_{N-j}.
\ee
$N_\k$ is not conserved; we have access to its thermal expectation value $\langle N_\k\rangle$ through the one-particle reduced density matrix, of which it is the eigenvalue~\cite{PO}. Indeed, for bosons
\bea
\langle N_\k\rangle&=&Z^{-1}\Tr \left[e^{-\beta H}N_\k\right]\nonumber\\
&=&\frac{N}{Z}\sum_{\k_1,\ldots,\k_N} \langle\k_1,\ldots,\k_N|N_{\k,1}e^{-\beta H}N_{\k,1}|\k_1,\ldots,\k_N\rangle\nonumber\\
&=&\frac{N}{Z}\sum_{\k_2,\ldots,\k_N} \langle\k,\k_2,\ldots,\k_N|e^{-\beta H}|\k,\k_2,\ldots,\k_N\rangle\nonumber\\
&\equiv&\langle\k|\rho_1|\k\rangle,
\eea
where
\be
\rho_1=NZ^{-1}\Tr_{2,\ldots,N}\,e^{-\beta H}
\ee
is the one-particle reduced density matrix. However, $\rho_1$ is diagonal in momentum representation: because $H$ conserves the total momentum,
\bea
\langle\k|\rho_1|\k'\rangle&=&\frac{N}{Z}\sum_{\k_2,\ldots,\k_N} \langle\k,\k_2,\ldots,\k_N|e^{-\beta H}|\k',\k_2,\ldots,\k_N\rangle \nonumber\\
&=&\delta_{\k,\k'}\langle\k|\rho_1|\k\rangle.
\eea
Therefore,
\be
\rho_1=\sum_{\k\in\Lambda^*}\langle N_\k\rangle |\k\rangle\langle\k|,
\ee
and its integral kernel is
\be
0\leq\langle\x|\rho_1|\y\rangle=L^{-d}\sum_{\k\in\Lambda^*}\langle N_\k\rangle e^{i\k\cdot(\x-\y)} \leq\langle\x|\rho_1|\x\rangle=\rho.
\ee
The kernel is nonnegative because it can also be obtained by integration from the integral kernel of $e^{-\beta H}$. Now
\be
0\leq n_\k\equiv\langle N_\k\rangle/N\leq 1,\quad \sum_{\k\in\Lambda^*}n_\k=1,
\ee
therefore $n_\k$ can be interpreted as the probability of the occurrence of a particle in the one-particle state $|\k\rangle$. We can also define
\be
\Gamma^1_{L}(\kappa)=\sum_{\k\in\Lambda^*:\|\k\|\leq\kappa}n_\k,
\ee
the probability distribution function for the single-particle momentum. Altogether, the conclusion is the same as for $\rho_{\rm c.m.}$.

\vspace{3pt}
\noindent
{\bf Proposition IV.1.} $\{Nn_\k,L^{-\frac{d}{2}}e^{i\k\cdot\x}\}_{\k\in\Lambda^*}$ are the eigenvalues and corresponding eigenfunctions of the one-particle reduced density matrix $\rho_1$. The integral kernel $\langle \bfz|\rho_1|\z\rangle$ is a nonnegative positive definite function of $\z$ on $\R^d$. Moreover, $n_\k$ is a positive definite function of $\k$ on $\Lambda^*$; in particular, $n_\bfz\geq n_\k$.

\vspace{3pt}
Thus, $\rho_1$ has the same structural properties as $\rho_{\rm c.m.}$, and $n_\k$ is the analogue of $\nu_\k$, $\k$ going over the same set $\Lambda^*$. However, their $\k$-dependence or the $\kappa$-dependence of $\Gamma^N_{L}$ and $\Gamma^1_{L}$ is expected to be quite different, owing to the fact that the dependence of $\psi_{\Q,n}(\x_1,\ldots,\x_N)$ on $\overline{\x}$ is separable while its dependence on $\x_j$ is not. In what follows, we investigate the possible alternatives, with special attention to the limit of infinite space.

\vspace{5pt}
\noindent
{\bf V. LIMIT OF SEQUENCES OF POSITIVE DEFINITE ATOMIC PROBABILITY MEASURES}

\vspace{3pt}
In both cases, that of the total and of the single-particle momentum, we are given a sequence of cubes $\Lambda=[-L/2,L/2]^d$ whose side length $L$ tends to infinity, the associated dual lattices $\Lambda^*=(2\pi/L)\Z^d$, and for each $L$ an even probability measure $\varphi_L$ on $\Lambda^*$:
\be\label{fiL}
\varphi_L(\k)=\varphi_L(-\k)\geq 0,\quad \sum_{\k\in\Lambda^*}\varphi_L(\k)=1.
\ee
Thus, the Fourier transform
\be\label{def-fL}
f_L(\x)=\sum_{\k\in\Lambda^*}\varphi_L(\k)e^{-i\k\cdot\x}
\ee
is a real even continuous positive definite $\Lambda$-periodic function on $\R^d$, $f_L(\x)\leq f_L(\bfz)=1$. Moreover, we know that $f_L(\x)\geq 0$. This implies that
\be\label{varfik}
\varphi_L(\k)=L^{-d}\int_\Lambda f_L(\x)e^{i\k\cdot\x}\d\x
\ee
is positive definite on $\Lambda^*$. In the present applications $\varphi_L(\k)$ is either $\nu_\k$ or $n_\k$ and, accordingly, $f_L(\x)$ is either $\rho^{-1}\langle\bfz|\rho_{\rm c.m.}|\x\rangle$ or $\rho^{-1}\langle\bfz|\rho_1|\x\rangle$.

For $\kappa\geq 0$ let
\be\label{GammaL}
\Gamma_L(\kappa)=\sum_{\k\in\Lambda^*, \|\k\|\leq\kappa}\varphi_L(\k)
\ee
and
\be\label{defGamma}
\Gamma(\kappa)=\limsup_{L\to\infty}\Gamma_{L}(\kappa).
\ee
The corresponding quantity for the total and single-particle momenta are $\Gamma^{\rm c.m.}$ and $\Gamma^1$, respectively.

\vspace{3pt}
\noindent
{\bf Proposition V.1.} $\Gamma(\kappa)$ is monotone increasing.

\vspace{3pt}
\noindent
{\em Proof.} Let $\kappa<\kappa'$, and let $L_i$ be a subsequence such that
$$
\lim_{i\to\infty}\Gamma_{L_i}(\kappa)=\Gamma(\kappa).
$$
Because $\Gamma_{L}$ is monotone increasing for any $L$,
$$
\Gamma_{L_i}(\kappa)\leq\Gamma_{L_i}(\kappa')
$$
for each $i$. By the Bolzano-Weierstrass theorem, there is a subsequence $L_{i_j}$ of $L_i$ on which convergence takes place at $\kappa'$, and necessarily
$$
\Gamma(\kappa')=\limsup_{L\to\infty}\Gamma_{L}(\kappa')
\geq\lim_{j\to\infty}\Gamma_{L_{i_j}}(\kappa') \geq \Gamma(\kappa).\quad\Box
$$

To discuss limits when $L$ goes to infinity, it is useful to extend $\varphi_L$ from $\Lambda^*$ to $(\R^d)^*$ by using Dirac deltas:
\be\label{extended}
\tilde{\varphi}_L=\sum_{\q\in\Lambda^*}\varphi_L(\q)\delta_\q,\quad \tilde{\varphi}_L(\k)=\sum_{\q\in\Lambda^*}\varphi_L(\q)\delta(\k-\q).
\ee
Then
\be\label{fL-extended}
f_L(\x)=\int \tilde{\varphi}_L(\k)e^{-i\k\cdot\x}\d\k
\ee
and, if $\kappa$ is not an integer multiple of $2\pi/L$ then
\be
\Gamma_L(\kappa)=\int_{\|\k\|\leq\kappa}\tilde{\varphi}_L(\k)\d\k.
\ee
If $\tilde{\varphi}_L$ converges to $\tilde{\varphi}$ in distribution sense then $\Gamma_L$ is convergent and
\be
\Gamma(\kappa)=\int_{\|\k\|\leq\kappa}\tilde{\varphi}(\k)\d\k.
\ee
This holds because the characteristic function of the cube $\{\|\k\|\leq\kappa\}$ can be approximated by functions of rapid decrease. However, $\tilde{\varphi}$ may not be a probability measure, because
\be\label{Gamma(k<infty)}
\Gamma_{<\infty}\equiv\lim_{\kappa\to\infty}\Gamma(\kappa)=\int \tilde{\varphi}(\k)\d\k
\ee
may be less than 1. We can remedy this by a one-point compactification of $(\R^d)^*$,
\be
\overline{(\R^d)^*}=(\R^d)^*\cup\{\infty\},
\ee
and defining
\be\label{Gamma-infty}
\Gamma_\infty=1-\Gamma_{<\infty}.
\ee
Let
\be
G_\lambda(\x)=(\pi\lambda^2)^{-d/2}e^{-\x^2/\lambda^2}.
\ee
When $\lambda$ goes to zero, $G_\lambda$ tends to $\delta_\bfz$ in distribution sense.

\vspace{5pt}
\noindent
{\bf Proposition V.2.} If $\tilde{\varphi}_L$ tends to $\tilde{\varphi}$ in distribution sense then
\be
\tilde{f}(\x):=\lim_{\lambda\to 0}\lim_{L\to\infty}(f_L*G_\lambda)(\x)=\int\d\k\,\tilde{\varphi}(\k)e^{-i\k\cdot\x},
\ee
implying the existence of the limit, and
\be\label{Gtildef}
\Gamma_{<\infty}=\tilde{f}(\bfz).
\ee
{\em Proof.}
\bea
(f_L*G_\lambda)(\x)&=&\int\d\k\,\tilde{\varphi}_L(\k)e^{-i\k\cdot\x}\int\d\y\, e^{i\k\cdot\y}G_\lambda(\y)
\nonumber\\
&=&\int\d\k\,\tilde{\varphi}_L(\k)e^{-i\k\cdot\x}e^{-\lambda^2\k^2/4}.
\eea
Because in the integrand $\tilde{\varphi}_L(\k)$ is multiplied with a function of rapid decrease, one has
\be
\lim_{L\to\infty}(f_L*G_\lambda)(\x)=\int\d\k\,\tilde{\varphi}(\k)e^{-i\k\cdot\x}e^{-\lambda^2\k^2/4}.
\ee
Setting $\lambda=0$ and $\x=\bfz$, comparison with Eq.~(\ref{Gamma(k<infty)}) yields the result. $\Box$

\vspace{3pt}
We define
\be
f(\x)=\lim_{L\to\infty}f_L(\x)
\ee
provided that the limit exists. If this is the case, the question naturally arises whether $f=\tilde{f}$. If $\Gamma_{\infty}>0$, the answer is not, because
\be
f(\bfz)-\tilde{f}(\bfz)= 1-\Gamma_{<\infty}=\Gamma_\infty.
\ee
However,
\be\label{f=tildef}
f(\x)=\tilde{f}(\x)\quad\mbox{if $\x\neq\bfz$}
\ee
may hold true. Actually, this can easily be seen in one dimension, because for $x\neq 0$, $\cos kx$ can be piecewise approximated by functions of rapid decrease. In the examples given below we shall assume Eq.~(\ref{f=tildef}) also for $d>1$. This implies that
\be\label{Gfto0}
\Gamma_{<\infty}=\lim_{\x\to \bfz}f(\x),
\ee
so that if $\Gamma_{<\infty}<1$ then $f(\x)$ is discontinuous at zero with a jump $\Gamma_\infty$.

$\Gamma_{<\infty}$ and $\Gamma_\infty$ can be interpreted as the probabilities that in the limit of infinite space $\k$ is finite or infinite, respectively. Here is a quick result about  the positivity of $\Gamma_\infty$.

\vspace{5pt}
\noindent
{\bf Proposition V.3.} Suppose that there is a sequence $\kappa_{L}$ such that $\lim_{L\to\infty}\kappa_L=\infty$ and
$$
\limsup_{L\to\infty}\Gamma_{L}(\kappa_{L})<1.
$$
Then $\Gamma_\infty>0$.

\vspace{2pt}
\noindent
{\em Proof.} For any $\kappa<\infty$, $\Gamma_{L}(\kappa)\leq\Gamma_{L}(\kappa_{L})$ if $L$ is large enough, and thus
$$
\Gamma(\kappa)\leq\limsup_{L\to\infty}\Gamma_{L}(\kappa_{L}).
$$
Because the right member is independent of $\kappa$, we can take the limit $\kappa\to\infty$ and obtain $\Gamma_{<\infty}<1$. $\Box$

\vspace{5pt}
\noindent
{\bf Corollary.}  In one dimension, applying the proposition to $\limsup_{L\to\infty}\Gamma^N_L(\kappa_L)$ with $\kappa_L=\rho\sqrt{\rho L}/a_L$ where $a_L\to\infty$, $a_L=o(\sqrt{L})$, from Eq.~(\ref{1D-aL}) we obtain that for all $\beta$ and $\rho$, $\Gamma^{\rm c.m.}_\infty=1$; in two dimensions, from Eq.~(\ref{low-2D-up}) with large enough $J$ it follows that $\Gamma^{\rm c.m.}_\infty>0$. Later we shall associate these findings with the absence of phase transitions in 1D and the absence of fluid-solid phase transition in 2D.

\vspace{5pt}
\noindent
{\bf Examples.} (i) Let $\b_1,\ldots,\b_d$ be linearly independent vectors and
\be\label{LL*}
\LL^*=\{2\pi n_1\b_1+\cdots+2\pi n_d\b_d\}_{n_1,\cdots,n_d\in\Z}.
\ee
Suppose that the extension of $\nu_\k$ to $(\R^d)^*$ tends in distribution sense to an atomic probability measure concentrated on $\LL^*$:
\be\label{crystal}
\tilde{\nu}_L=\sum_{\k\in\Lambda^*}\nu_{\k}\delta_\k\rightarrow
\tilde{\nu}=\sum_{\q\in\LL^*}\tilde{\nu}_{\q}\delta_\q.
\ee
Then the infinite space limit of the autocorrelation function of the center of mass,
\be\label{fcm}
f^{\rm c.m.}(\x)\equiv\lim_{L\to\infty}\rho^{-1}\langle\bfz|\rho_{\rm c.m.}|\x\rangle =\sum_{\q\in\LL^*}\tilde{\nu}_{\q}e^{-i\q\cdot\x}
\ee
is a continuous nonnegative $\LL$-periodic function where
$$\LL=\{n_1\a_1+\cdots +n_d\a_d\}_{n_1,\cdots,n_d\in\Z},\quad \a_i\cdot\b_j=\delta_{ij},$$
and
\be
\Gamma^{\rm c.m.}_{<\infty}=\sum_{\q\in\LL^*}\tilde{\nu}_{\q}=\lim_{\x\to\bfz}f^{\rm c.m.}(\x)=1.
\ee
We think this to be typical for crystals; see also the result of the paper~\cite{Suto} for the ground state in 1D and the forthcoming discussion in Section VII.

\noindent
(ii) Suppose that the extension of $n_\k$ to $(\R^d)^*$,
\be
\tilde{n}_L=\sum_{\k\in\Lambda^*}n_{\k}\delta_\k
\ee
tends to a normalized regular distribution $\tilde{n}$; thus,
$\int \tilde{n}(\k)\d\k=1.$
Then
\be
f^1(\x)\equiv\lim_{L\to\infty}\rho^{-1}\langle\bfz|\rho_1|\x\rangle =\int \tilde{n}(\k)e^{-i\k\cdot\x}\d\k,
\ee
a continuous nonnegative function decaying at infinity, and
\be
\Gamma^1_{<\infty}=\int \tilde{n}(\k)\d\k=\lim_{\x\to\bfz}f^1(\x)=1.
\ee
This is typical for the single-particle momentum distribution of bosons in the absence of Bose-Einstein condensation. If there is BEC, in the limit we have $\tilde{n}=\tilde{n}_\bfz\delta_\bfz+\tilde{n}_c$ with $\tilde{n}_\bfz>0$ and $\tilde{n}_c$ continuous, so that
$$f^1(\x)\to \tilde{n}_\bfz\quad\mbox{as $\x\to\infty$}.$$
The non-decay of $f^1$ is the sign of the off-diagonal long-range order~\cite{PO}. Periodic order can be seen also on the limit of $\tilde{n}_L$: if this contains a point measure at some $\k\neq\bfz$, in infinite space there is periodic order parallel to $\k$~\cite{PVZ,Suto2}. In all dimensions $\Gamma^1_{<\infty}=1$ always holds, otherwise the kinetic energy density would diverge in the thermodynamic limit~\cite{Suto2} .

\noindent
(iii) Suppose that $\tilde{\nu}_L$ tends to an unnormalized regular distribution $\tilde{\nu}$; thus,
$$0<\int\tilde{\nu}(\k)\d\k<1.$$
Then for any $\x\neq\bfz$,
\be
f^{\rm c.m.}(\x)\leq\lim_{\y\to\bfz}f^{\rm c.m.}(\y)
=\Gamma^{\rm c.m.}_{<\infty}
<1,
\ee
cf. Eqs.~(\ref{Gamma(k<infty)}) and (\ref{Gfto0}), while $f^{\rm c.m.}(\bfz)=1$. Thus, $f^{\rm c.m.}$ is discontinuous at zero, and the value of its jump is $\Gamma^{\rm c.m.}_\infty$.
In Section VII we shall argue that the situation described in this point is characteristic to fluid phases. Outside the origin $f^{\rm c.m.}$ is a continuous function which in ordinary fluids decays at infinity. A special case is
\be\label{superstate}
\tilde{\nu}=\tilde{\nu}_\bfz\delta_\bfz+\tilde{\nu}_c
\ee
with $\tilde{\nu}_\bfz>0$ and $\tilde{\nu}_c$ continuous, when
\be\label{fcm-general}
f^{\rm c.m.}(\x)=\int \tilde{\nu}(\k)e^{-i\k\cdot\x}\d\k\to\tilde{\nu}_\bfz\quad\mbox{as $\x\to\infty$}.
\ee
We shall associate this case with superfluids.

\vspace{3pt}
We now prepare some further results about $\Gamma_{<\infty}=1$ and $\Gamma_\infty=1$. Combining Eqs.~(\ref{varfik}) and (\ref{GammaL}),
\be\label{GammaDirichlet}
\Gamma_L\left(\frac{2\pi n}{L}\right)=L^{-d}\int_\Lambda f_L(\x)\prod_{i=1}^d D_n\left(\frac{2\pi x_i}{L}\right)\d\x,
\ee
where
\be
D_n(x)=\sum_{l=-n}^n e^{ilx}=\frac{\sin[(2n+1)x/2]}{\sin(x/2)}
\ee
is the Dirichlet kernel. A well-known fact is that
\be\label{norm-of-Dn}
\int_{-\pi}^\pi\left|D_n(x)\right|\d x=O(\log n).
\ee
We will also need the Fej\'er kernel,
\be
F_n(x)=\frac{1}{n}\sum_{l=0}^{n-1}D_l(x) = \frac{\sin^2(nx/2)}{n\sin^2(x/2)} =\frac{1-\cos (nx)}{n(1-\cos x)}.
\ee
$F_n(x)$ is nonnegative and
\be
\frac{1}{2\pi}\int_{-\pi}^\pi F_n(x)\d x=1.
\ee
The graph of $F_n$ consists of a sequence of peaks separated by zeros at $2\pi l/n$, $l=\pm 1,\pm 2,\ldots$. The following inequality estimates the distribution of the weight under the central peak.

\vspace{3pt}
\noindent
{\bf Lemma V.1.} For any $n\geq 2$,
\be\label{Fcentral}
\frac{1}{2\pi}\int_{-\alpha/n}^{\alpha/n}F_n(x)\d x\geq \left\{\begin{array}{lll}
\frac{4}{\sqrt{3}\pi},&\mbox{if}&\sqrt{12}\leq\alpha\leq 2\pi\\
\frac{\alpha}{\pi}\left(1-\frac{\alpha^2}{36}\right)&\mbox{if}& 0<\alpha\leq\sqrt{12}.
\end{array}\right.
\ee

\vspace{3pt}
\noindent
{\em Proof.} The result is obtained by integrating the lower bound
\be
F_n(x)\geq\max\left\{n\left(1-\frac{n^2x^2}{12}\right),0\right\}.\ \Box
\ee

\vspace{3pt}
Next, we estimate the number of peaks that carry a prescribed weight.

\vspace{3pt}
\noindent
{\bf Lemma V.2.} For any $\varepsilon>0$ there exists an integer $l_0$ such that for any $n\geq 2l_0$,
\be\label{Fejer}
\frac{1}{2\pi}\int_{-2\pi l_0/n}^{2\pi l_0/n}F_n(x)\d x\geq 1-\varepsilon.
\ee

\vspace{3pt}
\noindent
{\em Proof.} Let $l$ be any integer in the interval $[1,n/2]$. Then
\be
\frac{1}{2\pi}\int_{\frac{2\pi l}{n}}^{\frac{2\pi(l+1)}{n}}F_n(x)\d x
\leq\frac{1}{n^2\left[1-\cos\frac{2\pi l}{n}\right]} \leq\frac{1}{\pi\left[1-\frac{\pi^2}{12}\right]l^2}.
\ee
Here we used $1-\cos x\geq x^2/2-x^4/24$ and $l/n\leq 1/2$. Because $\sum_{l=1}^\infty l^{-2}$ is convergent, we can choose $l_0$ such that
\be\label{l0}
\frac{1}{\pi\left[1-\frac{\pi^2}{12}\right]}\sum_{l=l_0}^\infty l^{-2}\leq \varepsilon/2.
\ee
Thus, for any $n\geq 2l_0$,
\be
\frac{1}{2\pi}\int_{\frac{2\pi l_0}{n}}^\pi F_n(x)\d x\leq\varepsilon/2,
\ee
implying the result. $\Box$

\vspace{5pt}
\noindent
{\bf Lemma V.3.} Let ${\cal F}$ be a sequence of real equicontinuous and uniformly bounded functions in a domain $\cal D$ of $\R^d$. Define
$f^+(\x)=\limsup_{f\in\cal F}f(\x)$ and $f^-(\x)=\liminf_{f\in\cal F}f(\x)$. Then $f^\pm$ are continuous in $\cal D$.

\vspace{3pt}
\noindent
{\em Proof.} We give the proof for $f^+$. Fix an $\varepsilon>0$ and an $\x\in{\cal D}$. Because of equicontinuity, there exists a $\delta>0$ such that for any $f\in\cal F$, $|f(\x)-f(\y)|<\varepsilon$ if $|\x-\y|<\delta$. Let $\y$ be such a point of $\cal D$. Let $\{g_n\}$ and $\{h_n\}$ be two subsequences from $\cal F$ such that $g_n(\x)\to f^+(\x)$, $h_n(\y)\to f^+(\y)$ as $n\to\infty$. There exists some $N_0$ such that for any $n>N_0$,
$$|g_n(\x)-f^+(\x)|<\varepsilon,\quad |h_n(\y)-f^+(\y)|<\varepsilon,$$
$$g_n(\y)<f^+(\y)+\varepsilon,\quad h_n(\x)<f^+(\x)+\varepsilon,$$
the last two inequalities following from the definition of the limsup. On the other hand, $$|g_n(\x)-g_n(\y)|<\varepsilon, \quad |h_n(\x)-h_n(\y)|<\varepsilon.$$
Thus,
$$|g_n(\y)-f^+(\x)|<2\varepsilon,\quad |h_n(\x)-f^+(\y)|<2\varepsilon.$$
We may suppose that $f^+(\x)\leq f^+(\y)$. Then
$$
f^+(\y)-2\varepsilon< h_n(\x)<f^+(\x)+\varepsilon,
$$
that is,
$$
0\leq f^+(\y)-f^+(\x)<3\varepsilon.\quad \Box
$$

\vspace{5pt}
\noindent
We shall apply this lemma with ${\cal F}=\{f_L\}$. Now
\be
f^-(\x)=\liminf_{L\to\infty}f_L(\x),\quad f^+(\x)=\limsup_{L\to\infty}f_L(\x).
\ee
Because $0\leq f_L(\x)\leq f_L(\bfz)=1$, we have
\be
0\leq f^-(\x)\leq f^+(\x)\leq f^\pm(\bfz)=1.
\ee

\vspace{5pt}
\noindent
{\bf Theorem V.1.} Suppose that the family of functions $\{f_{L}\}$ is equicontinuous in a neighborhood $\cal D$ of the origin. Then $f^\pm(\x)$ is continuous in $\cal D$, so
$\lim_{\x\to\bfz} f^\pm(\x)=1.$ Moreover, $\Gamma_{<\infty}=1$, and for any sequence $\kappa_L$ going to infinity, $\Gamma_{L}(\kappa_{L})\to 1$.

\vspace{2pt}
\noindent
{\em Remarks.} 1. Instead of the equicontinuity of the whole sequence $f_L$, equicontinuity of some subsequence $f_{L_i}$ would suffice. 2. By the Arzela-Ascoli theorem, the sequence $\{f_{L}\}$ has a [not necessarily unique] continuous limit function in $\cal D$.

\vspace{5pt}
\noindent
{\em Proof.} The continuity of $f^\pm$ follows from Lemma 3. Equicontinuity implies that
$$f^-(\x)\leq\tilde{f}(\x)\leq f^+(\x)$$
where we can use limsup or liminf in the definition of $\tilde{f}$. This, together with Eq.~(\ref{Gtildef}) yields $\Gamma_{<\infty}=1$. Below we give also another proof for the saturation of $\Gamma$, that we shall use for the proof of $\Gamma_{L}(\kappa_{L})\to 1$. Define
\be
\sigma_{L}\left(\frac{2\pi n}{L}\right)=n^{-d}\sum_{l_1=0}^{n-1}\sum_{|k_1|\leq\frac{2\pi l_1}{L}}\cdots\sum_{l_d=0}^{n-1}\sum_{|k_d|\leq\frac{2\pi l_d}{L}}\varphi_L(\k),
\ee
then
\be\label{G>S}
\Gamma_{L}\left(\frac{2\pi n}{L}\right)\geq \sigma_{L}\left(\frac{2\pi n}{L}\right).
\ee
Indeed, $\Gamma_{L}(2\pi n/L)$ is obtained by summing $\varphi_L(\k)$ in the central cube of $\Lambda^*$ of side length $2\pi n/L$, while for $\sigma_{L}(2\pi n/L)$ we sum in centered rectangles inside this cube, and then average over the rectangles. Since all the terms are nonnegative and each rectangle contains only a part of those in the cube, the sums over the rectangles and, hence, also their average is smaller than the sum over the cube. Substituting $\varphi_L(\k)$ from Eq.~(\ref{varfik}) and summing inside the integral, after a change of variables we obtain
\be\label{sigmaL}
\sigma_{L}\left(\frac{2\pi n}{L}\right)=(2\pi)^{-d}\int_{\|\y\|<\pi} f_{L}\left(\frac{L}{2\pi}\y\right)\prod_{i=1}^d F_n(y_i)\d\y
\ee
where $\|\y\|=\max_i|y_i|$. Due to equicontinuity and $f_L(\bfz)=1$, if  $\varepsilon>0$ is given then for any small enough $\lambda$ and any $\x$ such that $\|\x\|\leq\lambda$, $f_{L}(\x)\geq 1-\varepsilon$. Consider only $L$ so large (or $\lambda$ so small) that $L/\lambda\geq 2$. Let $\eta=2\pi\lambda/L$, then
\be
\sigma_{L}\left(\frac{2\pi n}{L}\right)\geq (1-\varepsilon)\left[\frac{1}{2\pi}\int_{-\eta}^\eta F_n(x)\d x\right]^d.
\ee
Choose an integer $l_0$ such that the inequality~(\ref{l0}) holds true, and let $n=n(L,\varepsilon)=\lfloor l_0L/\lambda\rfloor\geq 2l_0$. Then
$$
\sigma_{L}\left(\frac{2\pi n(L,\varepsilon)}{L}\right)\geq (1-\varepsilon)^{d+1}.
$$
Now
$$\frac{2\pi n(L,\varepsilon)}{L}\leq\frac{2\pi l_0}{\lambda},$$
therefore
$$\Gamma\left(\frac{2\pi l_0}{\lambda}\right)\geq\lim_{L\to\infty}\sigma_{L}\left(\frac{2\pi l_0}{\lambda}\right)\geq (1-\varepsilon)^{d+1},$$
so that
\be\label{epsto0}
\lim_{\varepsilon\to 0}\Gamma\left(\frac{2\pi l_0}{\lambda}\right)=1.
\ee
Because $\Gamma$ is monotone increasing and bounded by 1 (or because $l_0/\lambda$ tends to infinity as $\varepsilon$ goes to zero), Eq.~(\ref{epsto0}) is equivalent to $\lim_{\kappa\to\infty}\Gamma(\kappa)=1$. Finally, consider any sequence $\kappa_L$ going to infinity and choose $n=\lfloor L\kappa_L/2\pi\rfloor$. Then $n>2l_0$ for $L$ large enough, and
$$
\lim_{L\to\infty}\Gamma_{L}(\kappa_L)\geq (1-\varepsilon)^{d+1}
$$
for any $\varepsilon>0$; thus, the left member must be 1. $\Box$

\vspace{3pt}
In what follows, we discuss the relation between $\Gamma_\infty=1$ and the properties of the sequence $f_L$.

\vspace{5pt}
\noindent
{\bf Proposition V.4.} If $\varphi_L(\bfz)=o\left(L^{-d}\right)$ then $\lim_{L\to\infty}f_L(\x)=0$ almost everywhere, and $\Gamma_\infty=1$.

\vspace{2pt}
\noindent
{\em Proof.} (i)
$$
\varphi_L(\bfz)=L^{-d}\int_\Lambda f_L(\x)\d\x=o(L^{-d})
$$
implies that $\int_\Lambda f_L(\x)\d\x=o(1)$. Then $\lim_{L\to\infty}f_L(\x)$ must vanish apart from a set of zero Lebesgue measure [which is nonempty since $f_L(\bfz)=1$ for each $L$].

\noindent
(ii) The number of lattice points in $\Lambda^*$ such that $\|\k\|\leq\kappa$ is not larger than $\left(\kappa L/\pi\right)^d.$ Because $\varphi_L(\k)\leq\varphi_L(\bfz)$, for any $\kappa<\infty$
$$
\Gamma_L(\kappa)\leq\varphi_L(\bfz)\left(\kappa L/\pi\right)^d\to 0\quad (L\to\infty)
$$
and, thus, $\Gamma_{<\infty}=0$.\ $\Box$

\vspace{5pt}
\noindent
{\bf Theorem V.2.} Suppose that there exists some $\kappa_0>0$ such that $\Gamma(\kappa_0)=0$. Then $\lim_{L\to\infty}f_L(\x)=0$ almost everywhere in $\R^d$.

\vspace{3pt}
\noindent
{\em Proof.} By the monotonic increase of $\Gamma$, $\Gamma(\kappa)=0$ for every $\kappa\leq\kappa_0$. Due to $\Gamma_L(\kappa)\geq\sigma_L(\kappa)$, c.f. Eq.~(\ref{G>S}),
$$
\limsup_{L\to\infty}\sigma_L(\kappa)=0\quad\mbox{if $\kappa\leq\kappa_0$.}
$$
For notational simplicity assume that $\kappa L/2\pi$ is an integer. Because the integrand in Eq.~(\ref{sigmaL}) is nonnegative,
\be\label{sigma-low}
\sigma_L(\kappa)\geq(2\pi)^{-d}\int_{\|\y\|<2\pi^2/\kappa L}f_{L}\left(\frac{L}{2\pi}\y\right)\prod_{i=1}^d F_{\kappa L/2\pi}(y_i)\d\y.
\ee
The boundary value $2\pi^2/\kappa L$ of the domain of integration corresponds to $\alpha=\pi$ in Eq.~(\ref{Fcentral}). It is easily seen that
$$
F_n(\pi/n)\geq 4n/\pi^2,
$$
and $F_n(x)$ increases monotonically when $|x|$ decreases from $\pi/n$ to zero, where its value is $n$. Thus, in the domain of integration $F_{\kappa L/2\pi}(y_i)$ varies within a factor 3. Applying Eq.~(\ref{Fcentral}) with $\alpha=\pi$, we obtain that
\be\label{F-low}
(2\pi)^{-d}\int_{\|\y\|<2\pi^2/\kappa L}\prod_{i=1}^d F_{\kappa L/2\pi}(y_i)\d\y\geq\left(1-\frac{\pi^2}{36}\right)^d.
\ee
In Eqs.~(\ref{sigma-low}) and (\ref{F-low}) we substitute $y_i=(2\pi/L)x_i$ and arrive at
\be
\sigma_L(\kappa)\geq \left(1-\frac{\pi^2}{36}\right)^d\frac{\int_{\|\x\|<\pi/\kappa}f_L(\x)\prod_{i=1}^d F_{\frac{\kappa L}{2\pi}}\left(\frac{2\pi x_i}{L}\right)\d\x}{\int_{\|\x\|<\pi/\kappa}\prod_{i=1}^d F_{\frac{\kappa L}{2\pi}}\left(\frac{2\pi x_i}{L}\right)\d\x}.
\ee
Fixing $\kappa\leq\kappa_0$ and letting $L$ go to infinity, $\sigma_L(\kappa)$ goes to zero, so $f_L(\x)$ must go to zero for almost every $\x$ in the cube $\{\|\x\|<\pi/\kappa\}$. Since this holds true for arbitrarily small $\kappa$, the result follows.\ $\Box$

Note that the condition of the above theorem is fulfilled if $\Gamma_\infty=1$. Therefore, we have the following.

\vspace{5pt}
\noindent
{\bf Corollary.} In one dimension $f^{\rm c.m.}(\bfz)=1$ and $f^{\rm c.m.}(\x)=0$ for almost all (probably for all) $\x\neq\bfz$.

Next, we prove a partial converse of Theorem V.2.

\vspace{5pt}
\noindent
{\bf Theorem V.3.} Suppose that $\lim_{L\to\infty}(\log L)^d f_L(\x)=0$ for almost every $\x\in\R^d$. Then $\Gamma_\infty=1$.

\vspace{3pt}
\noindent
{\em Proof.} From Eq.~(\ref{GammaDirichlet}), for $n=\kappa L/2\pi$,
\be
\Gamma_L(\kappa)\leq L^{-d}\int_\Lambda f_L(\x)\prod_{i=1}^d \left|D_{\kappa L/2\pi}\left(\frac{2\pi x_i}{L}\right)\right|\d\x.
\ee
By a variable transformation, Eq.~(\ref{norm-of-Dn}) is equivalent to
\be
L^{-1}\int_{-L/2}^{L/2}\left|D_{\kappa L/2\pi}\left(\frac{2\pi x}{L}\right)\right|\d x =O(\log\kappa L),
\ee
so there is some constant $C$ such that
\be
\Gamma_L(\kappa)\leq C(\log \kappa L)^d\ \frac{\int_{\Lambda}f_L(\x)\prod_{i=1}^d \left|D_{\frac{\kappa L}{2\pi}}\left(\frac{2\pi x_i}{L}\right)\right|\d\x}{\int_{\Lambda}\prod_{i=1}^d \left|D_{\frac{\kappa L}{2\pi}}\left(\frac{2\pi x_i}{L}\right)\right|\d\x}.
\ee
If $L$ goes to infinity, the right member goes to zero, so $\lim_{L\to\infty}\Gamma_L(\kappa)=0$ for every $\kappa$, which was the claim.\ $\Box$

\vspace{5pt}
\noindent
{\bf VI. FREE ENERGY DENSITY}

\vspace{3pt}

Because the upper bound in Eq.~(\ref{up-low}) is subexponential in $N$, the free energy density can be obtained from $Z_{\rm irred}$, c.f. Eq.~(\ref{Zirred0}). This we can rewrite as
\be\label{Zirred}
Z_{\rm irred}=e^{-\beta E_{\bfz,0}}\sum_{\q\in\Lambda^*_{\rm irred}}e^{-\beta\epsilon_\q}\,\Omega^\beta_N(\q),
\ee
where
\be\label{epsilon-q}
\epsilon_\q=E_{\q,0}-E_{\bfz,0}
\ee
and
\be\label{Omega}
\Omega^\beta_N(\q)=\sum_{n=0}^\infty e^{-\beta (E_{\q,n}-E_{\q,0})}.
\ee
Actually, it suffices to keep the $\q=\bfz$ term of $Z_{\rm irred}$. Let $\q_{\rm max}$ be the maximizer of $X^\beta_N(\q)=-\beta\epsilon_\q+\ln\Omega^\beta_N(\q)$. Because $\Lambda^*_{\rm irred}$ contains $N^d$ terms,
\be
e^{X^\beta_N(\q_{\rm max})} <e^{\beta E_{\bfz,0}}Z_{\rm irred}<N^d e^{X^\beta_N(\q_{\rm max})}.
\ee
The free energy density $f=-\beta^{-1}\lim_{L\to\infty} L^{-d}\ln Z$ will be given by
\be\label{f-from-qmax}
f= e_0-\beta^{-1}\lim_{L\to\infty} L^{-d} X^\beta_N(\q_{\rm max})
\ee
where $e_0=\lim L^{-d}E_{\bfz,0}$. From
\be
\nu_\q= Z^{-1}e^{-\beta E_{\bfz,0}+X^\beta_N(\q)}
\ee
and Eq.~(\ref{qmax=0}) it follows that $\q_{\rm max}=\bfz$ in the case of Boltzmann or Bose statistics~\cite{rem2}. We conclude that
\be
f= e_0-\beta^{-1}\lim_{L\to\infty} L^{-d}\ln \Omega^\beta_N(\bfz).
\ee

The free energy density is the Legendre transform of the microcanonical entropy density or, equivalently,
\be\label{Legendre}
f=e_0+\rho\min_{s\geq 0}\left\{e(s)-\beta^{-1} s\right\},
\ee
where $e(s)$ is the excitation energy (energy measured from $E_{\bfz,0}$) per particle at a (dimensionless) entropy per particle $s$ in the thermodynamic limit.

\noindent
{\bf Proposition VI.1.}
\be\label{0-not-0}
e(s)=e(\bfz,s),
\ee
where $e(\bfz,s)$ is the limit of the excitation energy per particle in the ensemble restricted to zero total momentum.

\vspace{3pt}
\noindent
{\em Proof.} We write
\be
\Omega^\beta_N(\q)=\int_{0}^\infty e^{-\beta E}\d\!\left[e^{S_\q^{L,N}(E)}\right]
=\int_0^\infty e^{-\beta E_{\q}^{L,N}(S)+S}\d S.
\ee
Here $e^{S_\q^{L,N}(E)}=1+\max\{n:E\geq E_{\q,n}-E_{\q,0}\}$, the microcanonical partition function in an ensemble where $\q$ is also fixed. The entropy $S_\q^{L,N}(E)$ increases with $E$, and its inverse $E_{\q}^{L,N}(S)$ increases with $S$.
Because the sum~(\ref{Omega}) converges for all $\beta>0$, $E_{\q}^{L,N}(S)$ must increase faster than linearly. Actually, if for a fixed $\q$, $S/N\to s$ as $N\to\infty$, then $E_{\q}^{L,N}(S)/N\to e(\q,s)$ which is a convex increasing function of $s$~\cite{Ru2}. Therefore, the maximum of $-\beta e(\q,s)+s$ is attained and
\be\label{Om-asymp}
\Omega^\beta_N(\q)=e^{N\max_{s\geq 0}\{-\beta e(\q,s)+s\}+o(N)}.
\ee
Combining Eqs.~(\ref{f-from-qmax})-(\ref{Legendre}) with Eq.~(\ref{Om-asymp}) yields Eq.~(\ref{0-not-0}).\ $\Box$

The above result implies that on the level of thermodynamic functions the canonical ensemble is equivalent to the microcanonical ensemble restricted to $\Q=\bfz$. Strong equivalence, meaning that the average value of local observables is the same in the two ensembles, holds very probably as well.

\vspace{5pt}
\noindent
{\bf VII. TOTAL MOMENTUM AND THERMODYNAMIC PHASES}

\vspace{3pt}
The examples given in Section V were meant to represent the distribution of the total momentum in different types of thermodynamic phases. Our point was that $\tilde{\nu}$ and
$$f^{\rm c.m.}(\x)=\lim_{L\to\infty}\rho^{-1}\langle\bfz|\rho_{\rm c.m.}|\x\rangle$$
are qualitatively different in fluid, solid, and superfluid phases. Below we formulate some detailed conjectures in this regard. The first two conjectures complete the theorems of Section II. Those results are valid both for bosons and fermions, with or without interaction. In noninteracting systems something stronger can be expected.

\vspace{5pt}
\noindent
{\bf Conjecture VII.1.} In two dimensional noninteracting systems for any $\beta$ and $\rho$ the asymptotic distribution of $\P/\sqrt{N}$ is normal.

\vspace{5pt}
In 3D systems of bosons we have to count with BEC. In the noninteracting case the fluctuations of $N_\bfz$ in the regime of condensation are of order $N^{2/3}$~\cite{BP}. Therefore, the following conjecture is slightly more challenging.

\vspace{3pt}
\noindent
{\bf Conjecture VII.2.} In 3D, for noninteracting bosons, at all temperatures and densities the asymptotic distribution of $\P/\sqrt{N-\langle N_\bfz\rangle}$ is normal.

The verification or falsification of these two conjectures is within the reach of the existing mathematical methods. The intuition behind them is that in the absence of interactions only an uncorrelated random motion of particles is possible and this must lead to a central limit theorem~\cite{rem4}.

In what follows we focus on the interacting case. All the conjectures below refer to quantum systems with strongly tempered interactions. Strong temperedness is a condition on the decay rate of the interaction, necessary in order that the potential energy can be made periodic and that we can use periodic boundary conditions~\cite{Ru2,Suto3}.

\vspace{5pt}
\noindent
{\bf Conjecture VII.3.} In interacting systems above one dimension, at any temperature and density $\Gamma^{\rm c.m.}_{<\infty}>0$. That is, in the infinite system the total momentum is finite with a nonvanishing probability in all the thermodynamic phases.

The existence of collective excitations that carry a finite momentum provides a strong experimental evidence supporting this conjecture. Such collective modes exist in all the thermodynamic phases; think about the density waves in gases and liquids, the lattice vibrations in crystals. Strictly speaking, these excitations do not manifest themselves in thermal equilibrium. To see them, the equilibrium -- specifically, the evenness [$\pm$ symmetry] of the momentum distribution -- must be broken, which is done by the experiment. 'Collective' means that a macroscopic number of particles contributes to the formation of each of these modes; thus, their wave vector is proportional to the total momentum which, therefore, must be finite with a positive probability. One dimension is an exception. Because in 1D the probability that the total momentum is finite tends to zero as the size of the system increases [$\Gamma^{\rm c.m.}_{<\infty}=0$], we arrive at the following conclusion.

\vspace{5pt}
\noindent
{\bf Proposition VII.1.} In one dimensional quantum systems with strongly tempered interactions, at any positive temperature the density waves disappear in the thermodynamic limit.

The restriction to $T>0$ is crucial. The ground state can be periodically ordered~\cite{Suto}, and there certainly exist low-lying excited states which correspond to its periodic modulations, i.e., to density waves. The absence of density waves at $T>0$ is the synonym of the absence of structural phase transitions.

In the subsequent discussion of thermodynamic phases the use of states in infinite volume will be helpful. These are positive normalized linear functionals over the algebra of quasi-local observables. Here we introduce them through a strict minimum of complication. We suppose that all the limits below exist. Let
\bea\label{state-decomp}
Z^L_{\leq\kappa} &=& \sum_{\|\Q\|\leq\kappa}\sum_{n=0}^\infty e^{-\beta E_{\Q,n}},\quad
Z^L_{>\kappa} = \sum_{\|\Q\|>\kappa}\sum_{n=0}^\infty e^{-\beta E_{\Q,n}},\nonumber\\
{\cal S}^L &=& Z^{-1}e^{-\beta H},\nonumber\\
{\cal S}_{\leq\kappa}^L &=& \left[Z^L_{\leq\kappa}\right]^{-1}\sum_{\|\Q\|\leq\kappa}\sum_{n=0}^\infty e^{-\beta E_{\Q,n}}\Pi_{\Q,n},\nonumber\\
S^L_{>\kappa} &=& \left[Z^L_{>\kappa}\right]^{-1}\sum_{\|\Q\|>\kappa}\sum_{n=0}^\infty e^{-\beta E_{\Q,n}}\Pi_{\Q,n}.
\eea
Then $Z=Z^L_{\leq\kappa}+Z^L_{>\kappa}$ and
\be\label{state-<>kappa}
{\cal S}^L=\frac{Z^L_{\leq\kappa}}{Z^L_{\leq\kappa}+Z^L_{>\kappa}}\ {\cal S}_{\leq\kappa}^L +\frac{Z^L_{>\kappa}}{Z^L_{\leq\kappa}+Z^L_{>\kappa}}S^L_{>\kappa},
\ee
which tends to
\be
{\cal S} = \Gamma^{\rm c.m.}(\kappa)\ {\cal S}_{\leq\kappa} +\left[1-\Gamma^{\rm c.m.}(\kappa)\right]S_{>\kappa},
\ee
where
\be
{\cal S} = \lim_{L\to\infty}{\cal S}^L,\quad{\cal S}_{\leq\kappa}=\lim_{L\to\infty}{\cal S}_{\leq\kappa}^L, \quad S_{>\kappa}=\lim_{L\to\infty}S^L_{>\kappa}.
\ee
These limits are taken on local observables. The decomposition can be done for any $\kappa$. Sending $\kappa$ to infinity, with
\be
{\cal S}_{<\infty}=\lim_{\kappa\to\infty}{\cal S}_{\leq\kappa},\quad S_{\infty}=\lim_{\kappa\to\infty}{\cal S}_{>\kappa}
\ee
we have
\be
{\cal S}=\Gamma^{\rm c.m.}_{<\infty}\ {\cal S}_{<\infty} +\Gamma^{\rm c.m.}_\infty S_{\infty}.
\ee
The first term can further be decomposed. It reads
\be
\Gamma^{\rm c.m.}_{<\infty}\ {\cal S}_{<\infty}=\int\tilde{\nu}(\k)\sigma(\k)\d\k,
\ee
where
\be
\sigma(\k)=\lim_{q\to 0}\lim_{L\to\infty}\frac{\sum_{\|\Q-\k\|\leq q}\sum_{n=0}^\infty e^{-\beta E_{\Q,n}}\Pi_{\Q,n}}{\sum_{\|\Q-\k\|\leq q}\sum_{n=0}^\infty e^{-\beta E_{\Q,n}}}.
\ee
The final form of ${\cal S}$ is
\be\label{fluid}
{\cal S}=\int\tilde{\nu}(\k)\sigma(\k)\d\k +\Gamma^{\rm c.m.}_\infty S_{\infty}.
\ee
In Eq.~(\ref{fluid}) the first term is composed of finite momentum states and represents the collective modes; the second term is composed of infinite momentum states and represents the few-particle excitations.

Few-particle motion is a coordinated motion of a small group of particles which evolves independently from the rest of the system. Rotons are thought to belong to this category~\cite{Fey2}. In 3D at high energies, in 2D at all energies such nearly separated small groups occur with a positive density in a non-vanishing fraction of eigenstates. The sum of the group momenta will typically be of the order of $\sqrt{N}$. There exist also excited states whose total momentum is finite and is the sum of a finite number of single- or few-particle momenta. However, at $T>0$ the statistical weight of these states becomes negligible with the increasing system size.

\vspace{5pt}
\noindent
{\bf Conjecture VII.4.: Gases and liquids.} Above one dimension, in the fluid phases both probabilities, that of having a finite or an infinite total momentum in infinite space, are nonzero, i.e., $0<\Gamma^{\rm c.m.}_{<\infty}<1.$ Density waves have a finite total momentum whose probability distribution is continuous. The infinite total momentum is the resultant of random single or few-particle motions which in finite volumes create a total momentum of the order of $\sqrt{N}$. Thus, in Eq.~(\ref{fluid}), $\tilde{\nu}$ is continuous and $\Gamma^{\rm c.m.}_\infty>0$.

The difference between gas and liquid is only quantitative. When cooling the system, it is possible to go around the critical point by continuously varying the density. When cooling at a constant [global] density, there will be a phase transition with phase separation: the liquid phase has a higher density and a higher probability of a finite total momentum than the coexisting gas phase.

\vspace{5pt}
\noindent
{\bf Conjecture VII.5.: Solids.} In an infinite solid phase the random few-particle motion vanishes almost surely, implying that the total momentum is finite with probability 1. In a crystal the probability distribution of the total momentum in infinite space is concentrated on a lattice, cf. Eq.~(\ref{crystal}). So $\Gamma^{\rm c.m.}_\infty=0$, $\tilde{\nu}=\sum_{\k\in\LL^*}\tilde{\nu}_\k\delta_\k$, $\int\tilde{\nu}(\k)\d\k=1$, and
\be
{\cal S}=\sum_{\k\in\LL^*}\tilde{\nu}_\k\sigma(\k).
\ee

In view of what we found in 2D, this conjecture excludes the solid phases in 2D at any positive temperature. Indeed, by  Theorem II.2 and Proposition V.3 we know that $\Gamma^{\rm c.m.}_{\infty}>0$. This, together with Conjectures VII.3 and 4 predicts that 2D systems exist only in fluid [gas, liquid or superfluid] phases. This conclusion is in accordance with the preservation of translation invariance in 2D particle systems in continuous space~\cite{Mer,FP1,FP2,Rich}. In addition, there is the interesting consequence that if the hexatic phase were to exist classically, quantum fluctuations would destroy it: Orientational long-range or even quasi-long-range order is incompatible with an isotropic random few-particle motion which implies exponentially decaying correlations.

A special kind of crystals are the so-called coherent quantum crystals, introduced in the papers~\cite{KN,N}; see also Ref.~\cite{Suto2}. They are thought to be somewhere between crystals and fluids, showing periodic long-range order, BEC and fluid properties (large kinetic energy) simultaneously. In our interpretation their simplest representative would be characterized by an asymptotic probability measure of the form
\be
\tilde{\nu}=\tilde{\nu}_\bfz\delta_\bfz+\tilde{\nu}_\q[\delta_\q+\delta_{-\q}] +\int\tilde{\nu}_c(\k)\d\k+\Gamma^{\rm c.m.}_\infty,
\ee
where $\q\neq\bfz$, $\tilde{\nu}_\bfz\geq\tilde{\nu}_\q>0$, and $\tilde{\nu}_\bfz+2\tilde{\nu}_\q<1$. Bosons interacting with an integrable pair potential whose Fourier transform has a negative minimum can be in such a state at very low temperatures; $\q$ would be close to the minimum.

\vspace{5pt}
\noindent
{\bf Conjecture VII.6.: Superfluids.} A superfluid is a fluid in which the asymptotic probability that $\Q=\bfz$ is nonvanishing. That is, $\lim_{\kappa\to 0}\Gamma^{\rm c.m.}(\kappa)=\tilde{\nu}_\bfz>0$.

Note that $\lim_{L\to\infty}\nu_\bfz$ may vanish while $\tilde{\nu}_\bfz>0$.
According to this definition, a positive $\tilde{\nu}_\bfz$ should appear in 2D systems of bosons [helium films, trapped Bose gases] and in 3D liquid $^4$He when the temperature decreases below the transition temperature $T_s$. We remark that $\tilde{\nu}_\bfz>0$ also in crystals [$\{\tilde{\nu}_\k\}_{\k\in\LL^*}$ is positive definite, therefore $\tilde{\nu}_\bfz\geq\tilde{\nu}_\k$]; the specificity of superfluids compared to crystals is that outside $\k=\bfz$ the asymptotic distribution $\tilde{\nu}$ is continuous and $\Gamma^{\rm c.m.}_\infty>0$. This implies that $f^{\rm c.m.}(\x)$ has a jump at $\bfz$ and tends to the constant $\tilde{\nu}_\bfz>0$ as $\x\to\infty$, without exhibiting any long-range order. So we have
\be
\tilde{\nu}_\bfz+\int\tilde{\nu}_c(\k)\d\k+\Gamma^{\rm c.m.}_\infty=1,
\ee
all the three terms are positive, and $\tilde{\nu}_c$ is a continuous measure, cf. Eq.~(\ref{superstate}). The two-fluid state is
\be\label{s-n-decomp}
{\cal S}=\tilde{\nu}_\bfz\ {\cal S}_s+[1-\tilde{\nu}_\bfz]\ {\cal S}_n,
\ee
the super and normal fluid states are
\bea\label{super-normal}
{\cal S}_s &=& \sigma(\bfz)\nonumber\\
{\cal S}_n &=& [1-\tilde{\nu}_\bfz]^{-1}\left[\int\tilde{\nu}_c(\k)\sigma(\k)\d\k +\Gamma^{\rm c.m.}_\infty{\cal S}_{\infty}\right].
\eea
Thus, the super and normal fluid components are not parts of the system but its alternative states that compose the infinite volume Gibbs state by a convex combination. Conjecture VII.2 implies that $\tilde{\nu}_\bfz=0$ in the noninteracting Bose gas which is, therefore, not a superfluid~\cite{rem4}.

This image of a superfluid is radically different in many respects from what the well-established theory offers to us. One point is particularly important: in Eq.~(\ref{s-n-decomp}) the superfluid state has the same temperature and the same energy and entropy density [cf. Proposition VI.1] as the normal fluid state, and not zero temperature and zero entropy density, as Landau suggested it~\cite{Lan}. This is in accordance with the fact that the $\lambda$-transition is an equilibrium phase transition. In the discussion of the dissipative flow below we shall return to this question.

The above equations describe the superfluid at rest. Now we consider the superfluid in motion. A symmetric $N$-particle wave function $\Phi$ describes a system in motion of velocity $\v$ with respect to its container $\Lambda$ if
\be\label{fivfi}
\langle\Phi|\p_j|\Phi\rangle=m\v\quad\mbox{(any $j$)}.
\ee
Because then $\langle\Phi|\p_j-m\v|\Phi\rangle=0$, in the comoving reference frame [the one moving together with the system] the single-particle momenta will be
$\p_j^{\v}=\p_j-m\v$. Choose $\v$ so that $m\v/\hbar\in\Lambda^*$. Since $\Lambda^*$ fills densely the whole space when $L\to\infty$, this is not an important restriction. Introduce
\be
H^{\v}=\frac{1}{2m}\sum_{j=1}^N (\p_j^{\v})^2+U_\Lambda,\quad \P^{\v}=\sum_j\p_j^{\v}.
\ee
Then $H^{\v}$ and $\P^{\v}$ are, respectively, the energy and the total momentum operators in the comoving frame~\cite{rem3}. $H^{\v}$ has the same eigenstates as $H$. Because $m\v/\hbar\in\Lambda^*$, the eigenvalues of $H^{\v}$ are also unchanged [so $\Tr e^{-\beta H^{\v}}=Z$], only their assignment is permuted:
\bea\label{HgPg}
H^{\v}\psi_{\Q+Nm\v/\hbar,n}&=&E_{\Q,n}\ \psi_{\Q+Nm\v/\hbar,n},\nonumber\\ \P^{\v}\psi_{\Q+Nm\v/\hbar,n}&=&\hbar\Q\ \psi_{\Q+Nm\v/\hbar,n}.
\eea
Moreover,
\be\label{Galilei2}
\psi_{\Q+Nm\v/\hbar,n}=e^{iN(m/\hbar)\v\cdot\overline{\x}}\psi_{\Q,n}.
\ee
Equations (\ref{HgPg}) and (\ref{Galilei2}) together with Eq.~(\ref{Galilei}) express Galilean invariance~\cite{Suto}. The density matrix reduced to the center of mass becomes
\be
\rho^\v_{\rm c.m.}=N\sum_{\Q\in\Lambda^*}\nu_\Q |\Q+Nm\v/\hbar\rangle\langle\Q+Nm\v/\hbar|,
\ee
and therefore
\be
\langle\bfz|\rho^\v_{\rm c.m.}|\x\rangle =e^{-\frac{iNm}{\hbar}\v\cdot\x}\langle\bfz|\rho_{\rm c.m.}|\x\rangle.
\ee
Its square root can play the role of a macroscopic wave function,
\be
\Psi(\x)=\sqrt{\langle\bfz|\rho^\v_{\rm c.m.}|\x\rangle}=e^{-\frac{iNm}{2\hbar}\v\cdot\x} \sqrt{\langle\bfz|\rho_{\rm c.m.}|\x\rangle}.
\ee
For large $N$ and large $\x$, $\Psi(\x)\sim e^{-\frac{iNm}{2\hbar}\v\cdot\x}\sqrt{\rho\tilde{\nu}_\bfz}$.
Because the function under the square root is nonnegative, by taking the positive square root the phase of $\Psi$ is
\be
{\rm Arg\,}\Psi(\x)=-\frac{Nm}{2\hbar}\v\cdot\x,
\ee
so that the velocity is obtained as
\be
\v=-\frac{2\hbar}{Nm}\frac{\partial}{\partial\x}{\rm Arg\,}\Psi(\x).
\ee
The construction (\ref{state-decomp})-(\ref{fluid}) of the states in infinite volume can be repeated by replacing $\Pi_{\Q,n}$ with $\Pi_{\Q+Nm\v/\hbar,n}$ but leaving $E_{\Q,n}$ unchanged. All the infinite-volume states will be labeled by an upper index $\v$. The infinite volume Gibbs state (\ref{fluid}) becomes
\be
{\cal S}^\v=\int\tilde{\nu}(\k)\sigma^\v(\k)\d\k +\Gamma^{\rm c.m.}_\infty S^\v_{\infty}.
\ee
The two-fluid state of the superfluid is
\be
{\cal S}^\v=\tilde{\nu}_\bfz\ {\cal S}^\v_s+[1-\tilde{\nu}_\bfz]\ {\cal S}^\v_n,
\ee
with
\bea
{\cal S}^\v_s &=& \sigma^\v(\bfz)\nonumber\\
{\cal S}^\v_n &=& [1-\tilde{\nu}_\bfz]^{-1}\left[\int\tilde{\nu}_c(\k)\sigma^\v(\k)\d\k +\Gamma^{\rm c.m.}_\infty{\cal S}^\v_{\infty}\right]\nonumber\\
\phantom{a}
\eea
and unchanged super and normal fluid fractions. In general, $\tilde{\nu}$ remains unchanged because it depends only on the energies. For the same reason, the free energy density of ${\cal S}^\v$ agrees with that of ${\cal S}$.

The situation described above corresponds to a hypothetic frictionless flow. In a realistic flow the velocity dependence of the density matrix is not Galilean invariant. Below we show how dissipation leads to a critical velocity. Imagine that a constant flow velocity $\v$ is maintained e.g. by moving the capillary with velocity $-\v$, and $v=|\v|$ is smaller than the critical velocity $v_{\rm cr}(T)$ that we are going to determine. Now $\v$ will be the velocity only in the superfluid state, $\v_s=\v$. In the normal fluid state friction on the walls and viscosity reduce the mean drift velocity to $\v_n$, $|\v_n|<v$. The accompanying loss of kinetic energy is converted into heat which partly is dissipated into the environment, partly raises the temperature in both the super and the normal fluid states. In effect, the temperature of the superfluid component cannot be lower than that of the normal component, otherwise the superfluid would immediately disappear! This point is important because it contradicts Landau's intuition~\cite{Lan}, therefore we spell it out. Suppose that in the superfluid state the temperature is $T$, in the normal fluid state it is $T'$, and $T<T'<T_s$. Then in finite volumes the density matrix~(\ref{state-<>kappa}) becomes
\bea
{\cal S}^L=\frac{Z^L_{\leq\kappa}(T)}{Z^L_{\leq\kappa}(T)+Z^L_{>\kappa}(T')}\ {\cal S}_{\leq\kappa}^L(T)\nonumber\\ +\frac{Z^L_{>\kappa}(T')}{Z^L_{\leq\kappa}(T)+Z^L_{>\kappa}(T')}S^L_{>\kappa}(T').
\eea
The superfluid/normal fluid ratio is the double limit of
\be
\frac{Z^L_{\leq\kappa}(T)}{Z^L_{>\kappa}(T')}=\frac{Z^L_{\leq\kappa}(T)}{Z^L_{>\kappa}(T)}\cdot \frac{Z^L_{>\kappa}(T)}{Z^L_{>\kappa}(T')}.
\ee
Applying $\lim_{\kappa\to 0}\lim_{L\to\infty}$ to this equation, the first fraction on the right remains positive because $T<T_s$. However, the second fraction is exponentially small in $N$ and, thus, goes to zero as $L$ goes to infinity. [In finite volumes the ratio of two partition functions of the same system taken at different temperatures is exponential in $N$.]

The velocity of the normal fluid is time dependent, $\v_n=\v_n(t)$. Supposing that $\v_n(0)=\v$, we can write it in the form
\be
\v_n=\alpha(t)\v, \quad\alpha(0)=1.
\ee
As long as $\v_n\neq \bfz$, friction is in action, therefore $\v_n$ decreases and the heat production continues. Let $T_t$ denote the temperature of the system at time $t$, $T_0=T$. Since only one degree of freedom, the single-particle momentum parallel to $\v$ is concerned, conservation of energy dictates
\be\label{TtT}
\frac{1}{2}k_BT_t-\frac{1}{2}k_BT=\eta \frac{mv^2}{2}[1-\alpha(t)^2],
\ee
where $\eta\leq 1$ is the efficiency, i.e., the fraction of the heat that raises the temperature. Now $\eta$ may also depend on $t$, but we suppose that it does not tend to zero, and neglect this dependence. In this case $\alpha(t)\to 0$ as $t\to\infty$. At time $t$ the state of the system is
\be
{\cal S}^{\v}_{\rm noneq}(T_t)=\tilde{\nu}_\bfz(T_t)\ {\cal S}_{s}^\v(T_t)+\left[1-\tilde{\nu}_\bfz(T_t)\right]{\cal S}_{n}^{\alpha(t)\v}(T_t).
\ee
The state is out of equilibrium, because the two terms are generated by different Hamiltonians, $H^\v$ and $H^{\alpha(t)\v}$, respectively. The temperature $T_t$ increases in time; if $v$ was small enough then
$
\lim_{t\to\infty}T_t=T_v<T_s,
$
and ${\cal S}^{\v}_{\rm noneq}(T_t)$ converges to a non-Gibbsian steady state whose normal component is at rest in $\Lambda$ while the superfluid flows with velocity $\v$:
\be\label{nonGibbs}
{\cal S}^{\v}_{\rm noneq}(T_v)=\tilde{\nu}_\bfz(T_v)\ {\cal S}_{s}^\v(T_v)+\left[1-\tilde{\nu}_\bfz(T_v)\right]{\cal S}_{n}(T_v).
\ee
From Eq.~(\ref{TtT}),
\be
T_v=T+\eta mv^2/k_B.
\ee
As $v$ increases, $T_v$ tends to $T_s$, $\tilde{\nu}_\bfz(T_v)$ goes to zero, and thermal equilibrium is restored: the system will be in the state ${\cal S}_{n}(T_s)$, at rest with respect to $\Lambda$. The critical velocity is obtained from the equation $T_{v_{\rm cr}}=T_s$ which yields
\be
v_{\rm cr}(T)=\sqrt{k_B(T_s-T)/(\eta m)}.
\ee
A possible definition of the ground state in infinite volume is $\lim_{T\to 0}{\cal S}(T)$. A critical velocity for this state is provided by $v_{\rm cr}(0)=\sqrt{k_BT_s/(\eta m)}$. For He II at saturated vapor pressure $T_s=2.17$ K, which gives $v_{\rm cr}(0)=(67/\sqrt{\eta})$ m/s $\geq$ 67 m/s. Because of the walls, in a capillary there is normal fluid and dissipation also at $T=0$, so the above discussion is relevant: the dissipation will heat the system to the temperature $T_v=\eta mv^2/k_B$.

The mechanism by which friction affects the moving normal fluid but not the superfluid can be understood qualitatively. In the moving system friction on the walls is a surface effect, internal friction (viscosity) is a bulk effect, their best description is via a spatially random uncorrelated perturbation. Such a perturbation  will not excite collective modes, it can influence only small separated groups of particles and excite them to a (macroscopically) higher energy. These groups then relax and emit incoherent radiation which partly leaves the system, partly is reabsorbed and heats thereby. Now ${\cal S}^\v_s$ must be an infinitely entangled state, with no separable small groups of particles in it. Separated small groups of particles can be found in a macroscopic number only in the eigenstates $\psi_{\Q+Nm\v/\hbar,n}=e^{iN(m/\hbar)\v\cdot\overline{\x}}\psi_{\Q,n}$ with $|\Q|\sim\sqrt{N}$, which give rise to the ${\cal S}^\v_\infty$ component of ${\cal S}^\v_n$.


Our definition of the superfluid fraction $\tilde{\nu}_\bfz$ is in obvious analogy with the condensate fraction $\tilde{n}_\bfz$. The connection between superfluidity and BEC is a subtle question. In two dimensions there is no BEC~\cite{Hoh,Wag,BM} but superfluidity does exist~\cite{Dal,KK}. On the other hand, even though only an increased peak height and not a sharp peak could be measured at $\k=\bfz$ in the single-particle momentum distribution~\cite{Gly}, there is today a consensus based on analytic arguments, experiment and numerics~\cite{PO,AZ,SS,Gly,Cep,Mor} that in bulk liquid helium the superfluid transition and BEC occur simultaneously. However, the condensate ratio saturates at less than 10\% while the whole liquid becomes superfluid as the temperature goes to zero. So one cannot simply identify the superfluid with the Bose-condensate: apart from the numerical mismatch, they are conceptually different, the condensate being a part of the system while the superfluid is, according to us, a state of it. According to the Bogoliubov-Landau theory, in three dimensions BEC acts as a catalyst that triggers the superfluid transition. However, it may well be the other way around: when two phenomena are simultaneous, it is difficult to decide, which one is the cause and which is the consequence. We prefer to consider the emergence of the $\Q=\bfz$ subspace as the primary event that causes the superfluid transition both in 2D and in 3D and, additionally, BEC in three dimensions. Among many other signs, this also shows that the mechanism of BEC for interacting bosons must be quite different from the saturation effect which is at the origin of the condensation in the three dimensional noninteracting Bose gas.

\vspace{5pt}
\noindent
{\bf VIII. COMMENT ON LANDAU'S CRITERION OF SUPERFLUIDITY}

The physical source of the critical velocity $v_{\rm cr}(T)$, derived in the preceding section for superfluids in a capillary, is quite different from that of the critical velocity in Landau's theory. By analyzing Landau's original publication~\cite{Lan}, we show that his argument about the condition of excitability of the superfluid ground state in a dissipative flow contains an
error, which consists in deriving the critical velocity from a frictionless flow. Indeed, he obtained the critical velocity through a mere Galilean boost which, by preserving the energy spectrum, does not lead to dissipation.

The details are as follows. In Section 4 of Ref.~\cite{Lan}, Landau investigated the stability of the superfluid ground state of helium liquid in a capillary against low-energy excitations that a flow of velocity $v$ can create by losing kinetic energy. His argument is based on the properties of $\epsilon_\Q=E_{\Q,0}-E_{\bfz,0}$, the energy gap to the lowest-lying state of momentum $\hbar\Q$, that we introduced in Eq.~(\ref{epsilon-q}). This quantity often appears in rigorous works about Bose systems, see e.g. Refs.~\cite{Lieb,Sei}. In He II, $\epsilon_\Q$ is measured up to about 4\AA$^{-1}$. Its qualitative features, known already to Landau, very probably do not depend on $T$, provided that $T<T_s$: the curve starts linearly, passes over a maximum and exhibits the famous roton minimum~\cite{CW,Grif,Leg}. If $\epsilon_\Q$ could be measured at $T=0$, it would certainly show the same features as in the measurement~\cite{CW}, done at 1.1 K. In our notations, Landau wrote down the equation
\be\label{Lan1941}
\epsilon_{\q+Nm\v/\hbar}=\epsilon_\q+\hbar\q\cdot\v+\frac{1}{2}Nm\v^2,
\ee
first with $\epsilon_\q\approx c\hbar|\q|$ near $\q=\bfz$ [Eq. (4,1)] and second, with $\epsilon_\q\approx\Delta+(2\mu)^{-1}\hbar^2(\q-\q_r)^2$ [Eq. (4,3)] near the roton minimum $\q_r$. He considered Eq.~(\ref{Lan1941}) as an equation of energy balance for the moving fluid, and said that for a loss of kinetic energy and accompanying excitation,
\be\label{Lanineq}
\epsilon_\q+\hbar\q\cdot\v<0\quad\mbox{or}\quad v>\frac{\epsilon_\q}{\hbar|\q|}\quad\mbox{for some $\q$}
\ee
must hold. In particular, a phonon of momentum $\hbar\q$ can be excited if
\be
c|\q|+\q\cdot\v<0,
\ee
and a roton can be excited if
\be
\Delta+(2\mu)^{-1}\hbar^2(\q-\q_r)^2-\hbar|\q|v<0.
\ee
This, however, is not true, these inequalities do not point to any distinguished value of $v$. If we add $E_{\bfz,0}$ to both sides of Eq.~(\ref{Lan1941}), we obtain
\be
E_{\q+Nm\v/\hbar,0}=E_{\q,0}+\hbar\q\cdot\v+\frac{1}{2}Nm\v^2,
\ee
which is Eq.~(\ref{Galilei}) with $n=0$ and $\k=m\v/\hbar$ if we suppose, as everywhere in this paper, that $m\v/\hbar\in\Lambda^*$. By Galilean invariance~\cite{Suto}, the set equalities
\bea
\{E_{\q,0}+\hbar\q\cdot\v+\frac{1}{2}Nm\v^2\}_{\q\in\Lambda^*}&=&\{E_{\q+Nm\v/\hbar,0}\}_{\q\in\Lambda^*} \nonumber\\
&=& \{E_{\q,0}\}_{\q\in\Lambda^*},
\eea
hold true: the first equality is term-by-term, the second is true
with a permutation of the elements, which changes as $\v$ varies. In particular, as seen from Eq.~(\ref{HgPg}), 
\be
\psi_{Nm\v/\hbar,0}=e^{i(Nm/\hbar)\v\cdot\overline{\x}}\psi_{\bfz,0},
\ee
the lowest-lying eigenstate that describes the system moving with velocity $\v$ with respect to $\Lambda$ [$\psi_{Nm\v/\hbar,0}$ satisfies Eq.~(\ref{fivfi})], and $\psi_{\bfz,0}$, the ground state at $\v=\bfz$, exchange energy: that of the former decreases by $Nm\v^2/2$ and becomes $E_{\bfz,0}$, that of the latter increases by $Nm\v^2/2$ and becomes $E_{\bfz,0}+Nm\v^2/2$. This occurs for any velocity such that $m\v/\hbar\in\Lambda^*$; the sound velocity or the slope of the straight line drawn to the roton minimum are not distinguished.

More generally, one observes that $\epsilon_\Q$ plays no role: the excitation energy supplied by the kinetic energy loss of the real flow [which is dissipative even at $T=0$ because of the walls of the capillary] is of order $N$, highly above $\epsilon_\Q$ from $|\Q|=0$ to $|\Q|\sim\sqrt{N}$. Thus, if the critical velocity should come from a comparison of $\epsilon_\Q$ with the available excitation energy, then it would be zero.
Note, however, that condition~(\ref{Lanineq}) is relevant when bulk He II is locally perturbed, by moving in it a small object with velocity $\v$.

$\epsilon_\Q$ is nonetheless good for one thing. At $T>0$, still below the transition temperature, the zero point of the $|\Q|\rightarrow\epsilon_\Q$ curve does not belong to the ground state, but to a state of zero total momentum whose energy is in the order of $N$ higher than that of the ground state. This state cannot be else than the superfluid state at the given temperature, providing an experimental support to our view, that the superfluid state is a state of zero total momentum.

\vspace{5pt}
\noindent
{\bf IX. SUMMARY}

In this paper we studied the probability distribution of the total momentum of quantum fluids at positive temperatures. Sections II-VI contain mathematical results about this quantity in finite volumes and in the limit of infinite space. The motivation behind this work is our conviction that the distribution of the total momentum is significantly different in different thermodynamic phases, and therefore can be used to characterize fluids, solids and superfluids. This conviction is expressed in the examples given in Section V and through the conjectures formulated in Section VII. The main assumption is that in interacting quantum systems above one dimension the total momentum remains finite with a nonvanishing probability in the thermodynamic limit; for a justification we referred to the existence of collective excitations. One may then envisage that with some probability the infinite system can be in a state of zero total momentum. This is precisely what we think to happen during crystallization and in the superfluid phase transition. The difference between the two cases is that in crystals the total momentum is finite and is distributed on a lattice while in a superfluid it can be infinite and otherwise its distribution is continuous outside the origin. A measurement can project the system into any component of the thermal equilibrium state which has a non-vanishing probability, such as the superfluid component or a state with a distribution over nonzero finite momenta representing phonons in normal fluids or in crystals. We suggested the use of the density matrix reduced to the center of mass for the definition of a macroscopic wave function, and commented on the effect of dissipation. Section VIII was devoted to a critical discussion of Landau's criterion of superfluidity. The relation between BEC and the superfluid transition remains to be clarified. The major unsolved problem is, however, to prove the phase transitions.

\vspace{3pt}
\noindent
{\bf Acknowledgment.} This work was supported by OTKA Grant No. K109577.


\vspace{3pt}

\begin{thebibliography}{99}
\bibitem{Rue}
D. Ruelle: Existence of a phase transition in a continuous classical system. Phys. Rev. Lett. \textbf{27}, 1040-1041 (1971).
\bibitem{LMP}
J. L. Lebowitz, A. E. Mazel and E. Presutti: Liquid-vapor phase transitions for systems with finite-range interactions. Phys. Rev. Lett. \textbf{80}, 4701 (1998);
J. Stat. Phys. \textbf{94}, 955-1025 (1999).
\bibitem{BLRW}
L. Bowen, R. Lyons, C. Radin, and P. Winkler: Fluid-solid transition in a hard-core system. Phys. Rev. Lett. \textbf{96}, 025701 (2006).
\bibitem{Su0}
A. S\"ut\H o: Total momentum and thermodynamic phases of quantum systems. arXiv:1504.06141.
\bibitem{Suto}
A. S\"ut\H o: Galilean invariance in confined quantum systems: Implications for spectral gaps, superfluid flow, and periodic order. Phys. Rev. Lett. \textbf{112}, 095301 (2014).
\bibitem{Lan}
L. D. Landau: The theory of superfluidity of Helium II. J. Phys. (Moscow) {\bf 5}, 71-90 (1941).
\bibitem{Lon}
F. London, {\em Superfluids, Vol. 2: Macroscopic Theory of Superfluid Helium} (Wiley, New York, 1954)
\bibitem{Tisza}
L. Tisza: La viscosit\'e de h\'elium liquide et la statistique de Bose-Einstein. C. P. Paris {\bf 207}, 1035-1186 (1938).
\bibitem{NP}
P. Nozi\`eres and D. Pines, {\em Theory of Quantum Liquids II: Superfluid Bose Liquids} (Addison-Wesley, Redwood City, 1990).
\bibitem{Grif}
A. Griffin, {\em Excitations in a Bose-Condensed Liquid} (Cambridge University Press, 1993).
\bibitem{Leg}
A. J. Leggett, {\em Quantum Liquids} (Oxford University Press, 2006).
\bibitem{Kag}
M. Yu. Kagan, {\em Modern Trends in Superconductivity and Superfluidity} (Springer-Verlag, 2013).
\bibitem{Kad}
L. P. Kadanoff: Slippery wave functions. J. Stat. Phys. {\bf 152}, 805-823 (2013).
\bibitem{Fey1}
R. P. Feynman: Atomic theory of the $\lambda$ transition in helium. Phys. Rev. {\bf 91}, 1291-1301 (1953).
\bibitem{Wre}
W. F. Wreszinski: Landau superfluids as non-equilibrium stationary states. J. Math. Phys. \textbf{56}, 011901 (2015).
\bibitem{Hoh}
P. C. Hohenberg: Existence of long-range order in one and two dimensions. Phys. Rev. {\bf 158}, 383-386 (1967).
\bibitem{Wag}
H. Wagner: Long-wavelength excitations and the Goldstone theorem in many-particle systems with "broken symmetries". Z. Physik {\bf 195}, 273-299 (1966).
\bibitem{BM}
M. Bouziane and Ph. A. Martin: Bogoliubov inequality for unbounded operators and the Bose gas. J. Math. Phys. {\bf 17}, 1848-1851 (1976).
\bibitem{Dal}
R. Desbuquois, L. Chomaz, T. Yefsah, J. Léonard, J. Beugnon, Ch. Weitenberg, and J. Dalibard: Superfluid behaviour of a two-dimensional Bose gas. Nature Physics \textbf{8}, 645–648 (2012)
\bibitem{KK}
B. Kim and Y. Kwon: Structural and superfluid properties of the 4He monolayer on a C28 molecule. J. Low Temp. Phys. \textbf{171}, 599–605 (2013).
\bibitem{PO}
O. Penrose and L. Onsager: Bose-Einstein condensation and liquid helium. Phys. Rev. {\bf 104}, 576-584 (1956).
\bibitem{AZ}
L. Aleksandrov, V. A. Zagrebnov, Zh. A. Kozlov, V. A. Parfenov, and V. B. Prieezhev: High energy neutron scattering and the Bose condensate in He II. Sov. Phys.-JETP \textbf{41}, 915-918 (1976).
\bibitem{SS}
W. M. Snow and P. E. Sokol: Density and temperature dependence of the momentum distribution in liquid He4. J. Low Temp. Phys. {\bf 101}, 881-928 (1995).
\bibitem{Gly}
H. R. Glyde, R. T. Azuah and W. G. Stirling: Condensate, momentum distribution, and final-state effects in liquid He4. Phys. Rev. B {\bf 62}, 14337-14349 (2000).
\bibitem{Cep}
D. M. Ceperly: Path integrals in the theory of condensed helium. Rev. Mod. Phys. {\bf 67}, 279-355 (1995).
\bibitem{Mor}
S. Moroni, G. Senatore and S. Fantoni: Momentum distribution in liquid helium. Phys. Rev. B {\bf 55}, 1040-1049 (1997).
\bibitem{BaPe}
G. Baym and C. J. Pethick: Landau critical velocity in weakly interacting Bose gases. Phys. Rev. A {\bf 86}, 023602 (2012).
\bibitem{Het}
The notion for pure states appears in B. Het\'enyi: Drude weight, Meissner weight, rotational inertia of bosonic superfluids: How are they distinguished? J. Phys. Soc. Jpn. {\bf 83}, 034711 (2014).
\bibitem{RS}
M. Reed and B. Simon, {\em Functional Analysis} (Academic Press, New York, 1980).
\bibitem{PVZ}
J. V. Pul\'e, A. F. Verbeure, and V. A. Zagrebnov: On nonhomogeneous Bose condensation. J. Math. Phys. \textbf{46}, 083301 (2005).
\bibitem{Suto2}
A. S\"ut\H o: A possible mechanism of concurring diagonal and off-diagonal long-range order for soft interactions. J. Math. Phys. {\bf 50}, 032107 (2009).
\bibitem{rem2}
For fermions $\nu_\q$ is also extremal at zero by symmetry, and this extremum is certainly a maximum.
\bibitem{Ru2}
D. Ruelle, {\em Statistical Mechanics} (W. A. Benjamin, New York, 1969) 
\bibitem{Lieb}
E. H. Lieb: Exact analysis of an interacting Bose gas. II. The excitation spectrum. Phys. Rev. {\bf 130}, 1616-1624 (1963).
\bibitem{Sei}
R. Seiringer: The excitation spectrum of weakly interacting bosons. Commun. Math. Phys. {\bf 306}, 565-578 (2011).
\bibitem{CW}
R. A. Cowley and A. D. B. Woods: Inelastic scattering of thermal neutrons from liquid helium. Canad. J. Phys. {\bf 49}, 177-200 (1971) Fig. 6.
\bibitem{Fig}
See also Fig. 1.3 in Ref.~\cite{Grif} or Fig. 3.6 in Ref.~\cite{Leg}.
\bibitem{BP}
E. Buffet and J. V. Pul\`e: Fluctuation properties of the imperfect Bose gas. J. Math. Phys. {\bf 24}, 1608-1616 (1983).
\bibitem{rem4}
The proof of a local central limit theorem for the total momentum of the non-interacting Bose gas, confirming Conjecture VII. 2, has been obtained and will be published separately.
\bibitem{Suto3}
A. S\"ut\H o: Ground state at high density. Commun. Math. Phys. \textbf{305}, 657-710 (2011).
\bibitem{Fey2}
R. P. Feynman: Atomic theory of the two-fluid model of liquid helium. Phys. Rev. {\bf 94}, 262-277 (1954).
\bibitem{Mer}
N. D. Mermin: Crystalline order in two dimensions. Phys.Rev. \textbf{176}, 250-254 (1968).
\bibitem{FP1}
J. Fr\"ohlich and C.-E. Pfister: On the absence of spontaneous symmetry breaking and of crystalline
ordering in two-dimensional systems. Commun. Math. Phys. {\bf 81}, 277-298 (1981).
\bibitem{FP2}
J. Fr\"ohlich and C.-E. Pfister: Absence of crystalline ordering in two dimensions. Comm. Math.
Phys. \textbf{104}, 697-700 (1986).
\bibitem{Rich}
T. Richthammer: Translation-invariance of two-dimensional Gibbsian point processes. Commun. Math. Phys. \textbf{274}, 81-122 (2007).
\bibitem{KN}
D. A. Kirzhnits and Yu. A. Nepomnyashchii: Coherent crystallization of quantum liquid. Sov. Phys. JETP \textbf{32}, 1191-1197 (1971).
\bibitem{N}
Yu. A. Nepomnyashchii: Coherent crystals with one-dimensional and cubic lattices. Theor. Math. Phys. \textbf{8}, 928-938 (1971).
\bibitem{rem3}
In Ref.~\cite{Suto} the "$\g$ frame" moves with a velocity $-\hbar\g/m$ with respect to $\Lambda$. To be in accordance with the present notation, in Ref.~\cite{Suto} one should replace $\g$ in the upper indices by $\v$, but substitute $\g=-m\v/\hbar$ in the formulas.
\bibitem{rem5}
In one dimension $\epsilon(Q)$ is a periodic function of period length $2\pi\rho$, see Ref.~\cite{Suto}.
\end{thebibliography}
\end{document}